\documentclass[twocolumn,twoside,paper,10pt]{spwlatemplate}

	\newif\ifpdf
	\ifx\pdfoutput\undefined
	\pdffalse 
	\else
	\pdfoutput=1 
	\pdftrue
	\fi
	\ifpdf
	\usepackage[pdftex]{graphicx}
	\else
	\usepackage{graphicx}
	\fi

	\ifpdf
	\DeclareGraphicsExtensions{.pdf, .jpg}
	\else
	\DeclareGraphicsExtensions{.eps, .jpg}
	\fi
	
\usepackage{times}  
\usepackage{graphicx}
\usepackage[]{natbib}
\begin{document}

\def\r{{\bf r}}
\title{\large \uppercase{Effect of inhomogeneous surface relaxivity, pore geometry and internal field gradient on NMR logging:
exact and perturbative theories and numerical investigations}}
\address{Schlumberger Doll Research, Cambridge, MA 02139, USA}
\author{Seungoh Ryu }
\date{June 21-24, 2009}                                           
\spwlacopyright{Copyright 2009, held jointly by the Society of Petrophysicists and Well Log Analysts (SPWLA) and the submitting authors.\\
This paper was prepared for presentation at the SPWLA $50^{\rm th}$ Annual Logging Symposium held in The Woodlands, Texas, United States, June 21-24, 2009.}
\lefthead{\small SPWLA $50^{\rm th}$ Annual Logging Symposium, June 21-24, 2009}
\righthead{\small SPWLA $50^{\rm th}$ Annual Logging Symposium, June 21-24, 2009}

\begin{abstract}
Nuclear magnetic resonance is widely used as a probe of pore geometry and fluid composition in well logging. 
One of the critical assumptions often made is that the diffusion of fluid molecules is sufficiently fast to warrant the condition for direct mapping between the surface-enhanced relaxation rate and the pore geometry.
In pores satisfying such a condition, but having a significant spatial variation of surface relaxivity ($\rho$), one can show that the one-to-one mapping may break down. The degree to which  the NMR logging interpretation is affected has not been systematically studied until now. Extra relaxation due to diffusion in internal field gradient is another example where spatially varying relaxation strength may obscure the direct relationship. Better understanding of their interplay may be exploited to our advantage.
In this work, we theoretically investigate the interplay between the pore geometry, internal field  and the inhomogeneous surface relaxivity. We develop a perturbative framework and compare its results with exact solutions obtained for a class of $\rho$ textures in a pore with simple geometry. Its effect is quantified for a wide range of diffusivity and/or $\rho$ strength. The result allows us to set the bounds for the change in the final slope of the relaxation curve and may serve as a useful guide for logging applications in real rocks with a wide range of pore sizes and fluid diffusivity. We further employ large scale numerical simulations to perform virtual experiments on more complex situations. Internal field and its gradient distributions were obtained and analyzed for up to $1.5^3 {\rm cm}^3$  based on tomograms of carbonate rocks. We find that the texture of $\rho$ based on the internal field gradient induces a small, but observable shift, compared to a random noise generated texture for which no shift is observed.
\end{abstract}

\section{Pore geometry, $\delta\rho$ and NMR logging}
There exist potential pitfalls in the way NMR logs are interpreted \citep{Kleinberg:1996p811}. 
While it is widely agreed that the method is robust for simple types of porous media, key assumptions for its successful application may become compromised progressively as their geometrical and  lithological properties become complex.
To be precise, there are three necessary assumptions for  the simple mapping between the so-called $T_2$- and the pore size-distributions to work: (1) The pores are  {\em practically} isolated or periodic so that diffusive coupling\citep{Cohen:1982p593,deGennes:1982p682,Zielinski:2002p769} among the pores may be neglected. (2) Within each pore, the so-called {\em fast diffusion criterion} is satisfied so that the relaxation is controlled by the weak surface relaxation strength, which will be represented as $\rho$ ($\rho_0$, if uniform) from now on, rather than by the diffusive flux. The condition is given in terms of the control parameter as $\kappa \equiv \rho_0 L / D \ll 1$ 
assuming a cubic pore of volume $V=L^3$ and diffusivity $D$ of the fluid.\citep{Brownstein:1979p779}
(3) The mapping is based on the assumption, often made without any quantitative justification for a given rock, that $\rho$ is uniform across the pore-grain interface.
Extensive investigations were made on the first and the second\citep{McCall:1991p637, Wilkinson:1991p641, Bergman:1995p866, Ryu:2001p705, Ryu:2009p500, Zielinski:2002p769, Grebenkov:2007p774} issues, but the third has received relatively scant attention\citep{Ryu:2008p531,Ryu:2009p753,Ryu:2009p500, Arns:2006p581,Valfouskaya:2006p549}. 
Unless one incorporates all these issues on an equal footing, it becomes difficult to gauge uncertainty in an NMR log interpretation. 

The aim of this paper is to investigate systematically the nature of their violation and seek quantitative bounds for their experimental signatures, should they occur. This is done by first considering a simple pore with a class of $\rho(\r)$ textures which allows exact solutions for nontrivial cases. For a realistic pore geometry, we derive the pore geometry from 3D tomograms\citep{Sheppard:2004p795} and 
run random-walk simulations \citep{Ryu:2009p753} under a set of controlled spatial profiles of $\rho$. Our focus is on gaining better understanding of  whether and how the intertwined $\rho$ texture and the pore geometry would make the issues acute or negligible. Careful experimental characterization of $\rho$ is an invaluable component toward ultimately increasing the utility of NMR logging for challenging environments and such effort is emerging.\citep{Keating:2007p846} Our theory makes quantitative analyses possible in such investigations as well as guiding log interpretation when such information is unavailable.

The evolution of the polarized proton spin density may be analyzed effectively using the eigenmode analysis of the underlying Helmholtz problem. \citep{Brownstein:1979p779, Ryu:2001p705,Grebenkov:2007p774} For continuity, we will employ the notational convention used in our recent paper. \citep{Ryu:2009p500} Within the framework, 
the evolution of the total polarization ${\cal M}(t)$ is viewed as a superposition of the eigenmodes $\{ \phi_p (\r)\}$, ($p=0, 1, \ldots$) each with relaxation rate $\lambda_p$. For $\kappa \ll 1,$ the slowest mode with $p=0$ dominates; viewing a porous medium as an ensemble of pores with a distribution $P(\lambda_0)$ of $\lambda_0$ values, one arrives at the backbone of current NMR log interpretation. 

\begin{figure}
  \includegraphics[width=3in]{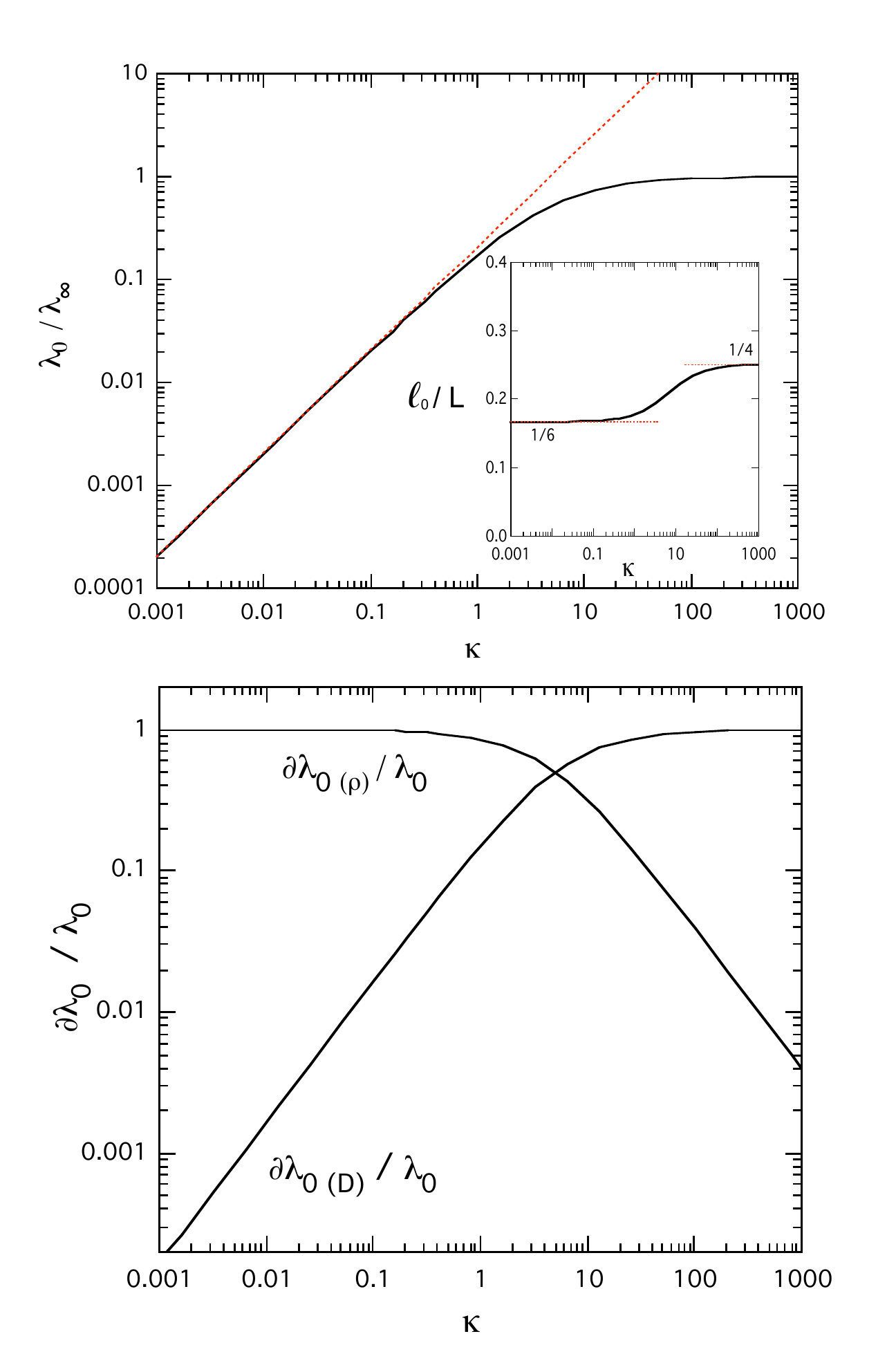} 
   \caption{\small Behavior of the slowest relaxation rate $\lambda_0$ as a function of the control parameter $\kappa \equiv \rho_0 L / D$ in a cubic pore of $V = L^3$ with uniform $\rho_0$. The inset to the first panel shows the effective pore size parameter $\ell_0 / L$. The value of $\ell_0/L = 1/6$ corresponds to the inverse surface-to-volume ratio of the pore. The bottom panel shows the relative fractional contributions to $\lambda_0$ from the surface-integral ($\partial \lambda_{0,(\rho)}$) and the volume-integral ($\partial \lambda_{0,(D)}$) contrbutions. (Eq.\ref{eq:decomposelambda})
}
\label{fig:basics2}
\end{figure}

Even though validity of the first two assumptions depends critically on the strength of $\rho_0$, its value, however, is often unknown.
A popular method to estimate the $\rho_0$ strength for a given porous medium is based on the assumption that the magnetization decays exponentially with its rate $\lambda_0 \sim \rho_0 S/V$. From an NMR probe-derived $T_2$ ($\sim 1/\lambda_0$) distribution, one may obtain an average $<\lambda_0>$, and 
combined with the $<S/V>$ value (independently measured such as from BET\citep{Kleinberg:1999p868,Keating:2007p846}), the estimated $\rho_0$ strength, $\tilde \rho_0$, may be obtained: $\tilde \rho_0 \sim <\lambda_0> / <S/V> $. To illustrate a potential pitfall in this approach, consider Figure \ref{fig:basics2} which shows how actual $\lambda_0$ varies as $\rho_0$ (and therefore $\kappa$) increases in a simple cubic pore of size $V= L^3$. While the linear relationship $\lambda_0 \propto \rho_0$ holds for $\kappa \ll 1$ as indicated by the borken line, the curve bends with an upper bound $\le \lambda_\infty \equiv D \frac{\pi^2}{L^2}$. This means that the {\em apparent} strength of $\rho_0$ as estimated by the above method, will then be upper-bounded,
\begin{equation}
\label{eq:apparentrho}
\tilde \rho_0 \le D \frac{\pi^2}{6 L}.
\end{equation}
For $L \sim 100 \mu m,$ and $D = 2500 \mu m^2/{\rm sec}$,  this means $\tilde \rho_0 \le 40 \mu m / {\rm sec}$.
As such an apparent $\rho_0$ value is bounded, the apparent $\kappa$ parameter
\begin{equation}
\label{eq:apparentkappa}
\tilde \kappa \sim \tilde\rho_0 \frac{L}{D} \le D \frac{\pi^2}{6 L} \frac{L}{D} =  \frac{\pi^2}{6} \sim {\cal O}(1)
\end{equation}
is also upper-bounded, which gives the erroneous impression that the premise $\kappa < 1$ is never grossly violated.
Using this circular reasoning, one would hardly encounter an empirical $\rho$ value that would suggest  the second condition is potentially violated. 
More accurate estimation of $\rho_0$ may be obtained if one uses the {\em initial} slope of the time-domain relaxation, which satisfies
\begin{equation}
\label{eq:taus}
\frac{1}{\tau_s} = \rho_0 \frac{S}{V}
\end{equation}
for all values of $\kappa$. However, the range over which it holds become extremely narrow as $\kappa$ increases, and therefore impractical for typical borehole applications. 
That leaves the rock-specific, accurate evaluation of $\rho_0$ still an unresolved issue and we should be mindful of the potential lapse in the {\em fast diffusion} condition when $\tilde \kappa \ge 0.1$ is observed. 

The extended nature of the pore space in natural media brings further issues. 
Suppose we manage to decompose such pore space into effectively isolated pore elements. 
How can one judge the robustness of the second assumption given these elements when their range of shapes and sizes vary significantly? 
For this, we first need to decide how to rephrase the second condition for general pore shape. 
The slowest relaxation rate $\lambda_0$ is composed of two components involving the slowest mode $\phi_0(\r)$,\citep{Ryu:2009p500}
\begin{equation}
\label{eq:decomposelambda}
\small
\lambda_0 = \oint d\sigma \phi_0^2(\r) \rho(\r) + D \int |\nabla \phi_0(\r)|^2 d\r \equiv \partial \lambda_{0,(\rho)} + \partial \lambda_{0,(D)}
\end{equation}
and we note that their relative fraction controls the degree to which the second condition is met. When $\partial \lambda_{0,(\rho)} \gg  \partial \lambda_{0,(D)}$, the relaxation is limited by the weak $\rho_0$, and is dominated by the slowest mode alone, validating the $T_2-$pore size mapping. In the opposite limit, the slowest mode profile acquires significant spatial variation, and higher modes gains weight, leading to a {\em multi-exponential} characteristic \citep{Brownstein:1979p779} which translates into a significant spread in the $T_2$ distribution. 
The bottom panel of Figure \ref{fig:basics2} shows how they behave over six decades in $\kappa (\equiv \rho_0 L / D)$ for a cubic pore. The crossover takes place over the range near $\kappa \sim 5$. For pores with a shape that defies characterization using a single length scale, the length scale defined as 
\begin{equation}
\label{eq:defineell0}
\ell_0 = \frac{\int d\r (\nabla \phi_0(\r))^2}{\oint d\sigma(\nabla \phi_0(\r))^2}
\end{equation} 
may be used for the $ \kappa_\ell \equiv \rho_0 \ell_0 / D$ parameter. 
It depends on the spatial profile of the slowest eigenmode and for cubic shape, it varies between $L/6$ (i.e. $V/S$) for $\kappa \sim 0$ and 
$L/4$ for $\kappa \rightarrow \infty$ as shown in the inset to the first panel of Figure \ref{fig:basics2}. In general, its asymptotic values in both $\kappa\rightarrow 0$ and $\rightarrow \infty$ limits should be characteristic of the pore shape.
Using this definition, we can put $\lambda_0$ into 
\begin{equation}
\label{eq:lam0alt}
\lambda_0 = \partial \lambda_{0,(\rho)} ( 1 + \rho_0 \ell_0 /D ),
\end{equation}
an expression we find useful later (Eq. \ref{eq:lam00quad}).

For a collection of isolated pores with varying shape, one would proceed to apply the {\em fast diffusion} criterion via $\rho_0 \ell_0 / D \ll 1$ for each individual pore. Isolated, large vugs, if present,  may obviously incur exceptions to the direct mapping. There are subtle effects beyond this in realistic porous media. 
Significant spatial fluctuations present in the local diffusive coupling strength \citep{Zielinski:2002p769} lead to enhanced inhomogeneity in $\phi_0(\r)$. This means that $\partial \lambda_{0,(D)}$ of Eq.\ref{eq:lam0alt} gains more weight, $\ell_0$ gets bigger, and the multi-exponential characteristics of the relaxation more pronounced.  As a result, the simple relationship between $\lambda_0$ and local pore geometry becomes obscure. When the first and the second conditions break down simultaneously, the combined effect is of pure geometrical nature, and it justifiably invites active current research on issues such as NMR log-prediction of permeability.  

Unfortunately, the difficulty may not stop at the purely geometrical level. Suppose we have a system with a uniform $\rho_0$ satisfying the first two criteria, for a range of pore sizes $a  \in [a_{min}, a_{max}]$. Then the observed rate distribution $P_{rate}(\lambda_0)$ leads directly to the size distribution $P_{size}(a)$. 
It is hypothetically possible, however, to have a collection of isolated pores of a suitably chosen size $a_0$, with a distribution $P_{\rho}(\rho_0)$ of $\rho_0$ values assigned to each,  that will yield the identical relaxation behavior. Since there exists scanty empirical data on $\rho(\r)$, and it is not always possible to have the actual 3D pore geometry known, it is not clear how to quantify {\em uncertainty} originating from break down of these assumptions respectively.
Accuracy of any physical property dependent on $\lambda_0$ and its probability distribution, such as permeability and porosity, will then be controlled by the uncertainty present in $\lambda_0$. 

To bring some clarification to this, we recently developed a theoretical method to solve for  the change in the relaxation rates $\{\lambda_p\}$ and their associated eigenmodes $\{ \phi_p \}$  which sets the bound for uncertainty for a property derived from $\lambda_p$'s.\citep{Ryu:2008p531,Ryu:2009p500}
In summary, we consider the spatially fluctuating part of the $\rho(\r)$ given by  
\begin{equation}
\label{eq:definedeltarho}
\delta \rho (\r) \equiv  \rho(\r)-\rho_0
\end{equation}
where $\rho_0$ is the average of $\rho(\r)$ over the interface, and derive the fractional shift in the slowest rate taking $\delta \rho/\rho_0$ as the perturbation,
\begin{eqnarray}
\label{eq:fractionalshiftsmalldeltarho2}
\frac{\delta \lambda_0}{\lambda_0^0} &\sim& \frac{ \delta\rho_{00}}{\lambda_0^0} \frac{S}{V} -  
 \sum_{q\ne 0} \frac{\delta\rho_{0q}\delta\rho_{q0}}{\lambda_0^0(\lambda_q^0-\lambda_0^0)} (\frac{S}{V})^2 
\end{eqnarray}
in terms of various surface overlap integrals that involve the eigenmodes $\phi_p^0$ of the uniform $\rho_0$ case, 
\begin{equation}
\delta\rho_{q0} \frac{S}{V}= \oint d\sigma \phi_0^0(\r) \delta \rho (\r) \phi_q^0(\r).
\end{equation}
In the case of a discrete hemispherical $\delta\rho$,\citep{Ryu:2009p500} an exact solution was also obtained, extending the bound for cases where $\delta \rho/\rho_0$ is not small. 
Due to high symmetry of the spherical pore, it left room for other simple cases where deformations in pore geometry and $\delta \rho$ patterns are more explicitly intertwined. 

In the following, we therefore extend the results to a rectangular pore.  A set of non-trivial  $\delta \rho$'s, appropriate for setting the bounds, are considered. 
For a rule-of-thumb type bound for log interpretation, we lay out steps to estimate the degree of change in $\delta \lambda_0 / \lambda_0$ (i.e. $\delta T_2/T_2$) expected for a combined geometrical and $\rho-$textural deviations from a pristine condition assuming that the porosity is preserved. 
We also demonstrate by numerical simulations the importance of relationship between the $\delta \rho (\r)$ texture and the $\phi_0(\r)$ profile and their symmetry. The manifestation of such effect is first considered for quadrature patterns of $\delta \rho$ in a cubic pore followed by cases with tomogram-derived 3D pore geometry. In the latter,  a $\delta \rho$ texture based on the internal field 
and another based on the correlated random noise sequence were imposed on the interface of a carbonate pore matrix. 

\section{Rectangular pore : Bounds for  $\rho(\r)$ and geometrical variations }
\label{sec:rect}
\subsection{{\it Defining pore geometry and $\rho$ textures}}
Once we allow general variations in $\delta \rho$ as well as pore geometry, the phase space quickly expands beyond our means. 
To make the problem tractable, yet meaningful, we consider a situation in which both vary under constraint in a space of dimension $d_e$. 
Consider a rectangular pore with the porosity function:
\begin{equation}
\phi(\r) = \Big\{ 
   \begin{array}{cl}
      1 & {\rm if\, } \/  r_\alpha  \in [-\frac{L_\alpha}{2} , \frac{L_\alpha}{2} ] {\rm \, for\, all\, }  \alpha  = 1, \ldots d_e \\
      0 & {\rm otherwise} \\
   \end{array}
\end{equation} 
with its geometry tuned by aspect ratios $L_\alpha / L_\beta$ preserving the total volume $V = \prod L_\alpha$. 
Such a variation fails to incorporate the {\em heterogeneity} present in natural media, but it is a good starting point. We systematiclly explore how $\delta \rho$ interferes with the pore geometry, specifically a porosity-preserving distortion and eventually a  {\em dimensional crossover}, and to what degree they affect the $T_2-$distribution. 

For the $\rho$ texture, we consider a situation where each side of the rectangle may have a distinct value $\rho_{\alpha\pm}$ where $\alpha\pm $  indicates the left ($-$) and the right ($+$) interface in the $\alpha-$direction . 
For each dimension, we introduce their average for each direction:
\begin{equation}
\bar{\rho}_\alpha \equiv \frac{\rho_{\alpha+} + \rho_{\alpha-}}{2}
\end{equation}
and the asymmetric variance
\begin{equation}
\quad \delta{\rho}_\alpha \equiv \frac{\rho_{\alpha+} - \rho_{\alpha-}}{2}.
\end{equation}
Generally, $\bar \rho_\alpha$ varies for each direction $\alpha$. 
Define gross average of $\bar\rho_\alpha$'s across the whole interface of the pore as 
\begin{equation}
\label{eq:globalrho}
<\bar \rho> = \frac{\sum_\alpha 2 \bar \rho_\alpha S_\alpha} {\sum_\alpha 2 S_\alpha}
\end{equation} where  $S_\alpha \equiv \prod_{\beta\ne \alpha}^{d_e} L_\beta$ is the cross-sectional area normal to $\hat{n}_\alpha$.
It is more convenient to use dimensionless control parameters,  $\sigma_\alpha$ and $\epsilon_\alpha$'s.
$\epsilon_\alpha$ defined as 
\begin{equation}
\label{eq:defineepsilon}
\epsilon_\alpha = \frac{\bar \rho_\alpha - <\bar \rho>}{ < \bar \rho> }
\end{equation}
is the relative deviation of each planar value $\bar\rho_\alpha$ from the average. 
$\epsilon_\alpha$ may vary for each direction, but they constitute {\em symmetric} part of the $\rho$ fluctuation as 
$\bar\rho_{\alpha\pm}$'s are equal in opposing planes unless $\sigma_\alpha \ne 0$,
where 
\begin{equation}
\label{eq:definesigmaalpha}
\sigma_\alpha \equiv \frac{\delta\rho_{\alpha}}{\bar{\rho}_\alpha}
\end{equation}
represents the {\em asymmetric} fluctuation of the $\rho$ between opposing planes.
Note that 
\begin{equation}
\label{eq:conditionforepsilon}
\sum_\alpha S_\alpha \epsilon_\alpha =V \sum_\alpha \frac{1}{L_\alpha} \epsilon_\alpha= 0
\end{equation}
by definition. 
Note that $\epsilon_\alpha$ and $\sigma_\alpha$ thus defined are analogous to the {\em compression/dialational} and the {\em shear} deformations in the theory of elastic deformation. 
Finally, we define
\begin{equation}
\label{eq:definekappaalpha}
\kappa_\alpha \equiv \frac{\bar\rho_\alpha L_\alpha}{D}
\end{equation}
as an analog in each dimension to the $\kappa$ parameter.

\begin{figure}[htbp] 
   \centering
   \includegraphics[width=3.3in]{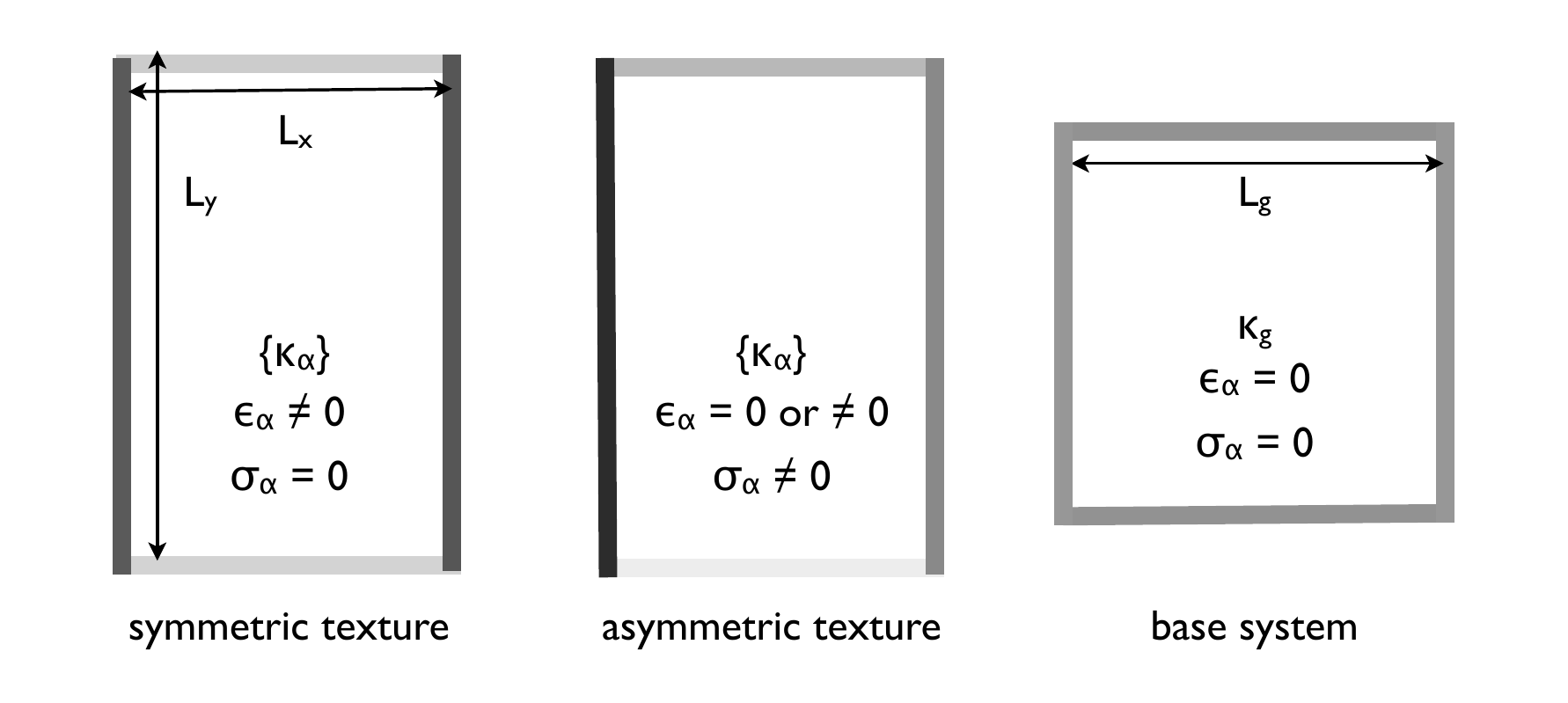} 
   \caption{\small Schematics of three cares of a rectangular pore. For clarity, 2-dimensional cross-sections are shown. The strength of the gray scale on the pore wall indicates the strength of $\rho(\r)$ on each plane. Left panel shows the {\em symmetric} rectangle (center panel) with $\sigma_\alpha = 0$ (for all $\alpha = x,y,z$); the center shows an {\em asymmetric} rectangle with $\sigma_\alpha \ne 0$. 
The right panel shows the base system, in which all $\bar\rho_\alpha$'s are equal to $\bar\rho_g = <\bar\rho>$ and $\epsilon_\alpha, \sigma_\alpha = 0$ in all directions. The volume of the base cube is chosen to match that of the rectangle.}
\label{fig:rectangles}
\end{figure}

We obtain analytic solutions for the slowest relaxation mode $\phi_0$ and its eigenvalue $\lambda_0$ for each of the three categories schematically described in Figure \ref{fig:rectangles}. The case with $\sigma_\alpha = 0$ for all $\alpha$ will be referred to as the {\em symmetric texture} and will be indicated by super/subscript $s$;  Cases with non-zero $\sigma_\alpha$ values, (but with $\epsilon_\alpha$ may or may not equal $0$) will be called the {\em asymmetric textured} with super/subscript $a$ where appropriate.
The globally uniform case is a special instance of the symmetric texture for which $\bar\rho_\alpha$ are all equal and will be indicated by super/subscript $g$ . 
With the symmetric texture, there is no {\em finite} minimum separation between a pair of interfacial points with distinct values of $\bar\rho$'s. Due to the separability of the coordinates, the solution for such situation can be trivially constructed out of the one-dimensional problem with the same boundary condition at both ends. 
We also impose
\begin{equation}
\bar\rho_\alpha \ge 0; \quad 1 \ge \sigma_{\alpha} \ge 0; 
\end{equation}
without losing generality. 
For symmetric textures, it is further required that $\sigma_\alpha = 0$ for all $\alpha$. The globally uniform case requires also that $\epsilon_\alpha =0$  for all $\alpha$.  $\sigma_\alpha = 1$ is rather special in that it forces the $\bar\rho_{\alpha-}$ to vanish on one of the opposing planes with an interesting consequence. (see Figure \ref{fig:ProfileForVariousSigmaAndKappas} and the discussion following it) 

To incorporate variations finer than described above, one may divide the rectangle into a set of smaller ones. The minumum size for these sub-rectangles will be set by the smallest length scale dictated by that of $\delta \rho(\r)$. Each of these sub-rectangles couples with its adjacent neighbors through the diffusive coupling, leading to a large set of coupled equations instead of a simple 2 by 2 homogeneous equation that we have here. 
One may also consider a sphere (or a cylinder) or ellipsoidal geometry. We have worked out a perturbative scheme for the spherical geometry\citep{Ryu:2009p500} and obtained exact solution for a simple binary hemispherical texture of $\rho$ \citep{Ryu:2009p753}. Extension  for more complex angular texture of $\delta \rho$ is straightforward in theory, but involves a large number of angular modes.
Rectangular (or cube) geometry makes the numerics much simpler for the class of $\rho$ textures of Figure \ref{fig:rectangles} while at the same time allowing a nontrivial geometrical variation. Note that the symmetric texture for a rectangle (or cube) corresponds to the $L=2 m$ ($m=1,2,\ldots$) spherical harmonics modes, while the asymmetric texture would bring in odd harmonics, $L=2 m + 1$, of the spherical pore.

\subsection{{\it General Solution}}
 The boundary condition for the general $\rho$ texture defined in Figure \ref{fig:rectangles} now leads to the pair of conditions in each dimension $\alpha$:
 \begin{eqnarray}
 \label{eq:rectbc}
&& \rho_{\alpha+} \phi_{0,\alpha}  (L_{\alpha}/2) + D \nabla_\alpha \phi_{0,\alpha}  (L_{\alpha}/2) = 0 \\
&&  \rho_{\alpha-} \phi_{0,\alpha}  (-L_{\alpha}/2) - D \nabla_\alpha \phi_{0,\alpha}  (-L_{\alpha}/2) = 0  \nonumber 
 \end{eqnarray}
 for the lowest eigenmode $\phi_0 (\{ r_\alpha \} )  = \prod_{\alpha}^{d_e} \phi_{0,\alpha} (r_\alpha) $ with each $\phi_{0,\alpha} $ in the following form:
 \begin{equation}
 \label{eq:groundstate}
 \phi_{0,\alpha} (r_\alpha) = A_\alpha \cos (k_\alpha r_\alpha) + B_\alpha \sin (k_\alpha r_\alpha)
 \end{equation}
with the normalization condition
\begin{equation}
\label{eq:normalize}
\small
A_{\alpha}^2  \int_{-\frac{L_\alpha}{2}}^{\frac{L_\alpha}{2}} dr_\alpha \cos^2 (k_\alpha r_\alpha) + 
B_{\alpha}^2 \int_{-\frac{L_\alpha}{2}}^{\frac{L_\alpha}{2}} dr_\alpha \cos^2 (k_\alpha r_\alpha) 
 = 1
\end{equation}
The constrained $\delta \rho$ textures and the geometry render the problem separable in each dimension, so we have a one-dimensional problem for each $\alpha$ satisfying: 
\begin{equation}
 {\cal K}_\alpha \cdot \Big(   \begin{array}{c}
      A_\alpha \\
      B_\alpha \\
   \end{array}
   \Big)   = 0
\end{equation}
with the matrix ${\cal K}_\alpha$ given by
\begin{equation}
{\small
\Big(  \begin{array}{cc}
      \sigma_\alpha \cos \frac{q_\alpha \pi}{2} & \frac{\pi}{\kappa_\alpha} q_\alpha \cos (\frac{q_\alpha \pi}{2})  + \sin  (\frac{q_\alpha \pi}{2})  \\
      \cos  \frac{q_\alpha \pi}{2} - \frac{\pi}{\kappa_\alpha} q_\alpha \sin \frac{q_\alpha \pi}{2}  & \sigma_\alpha \sin \frac{q_\alpha \pi}{2} \\
   \end{array}
   \Big) 
   }
   \end{equation}
and the dimensionless wavevector $q_\alpha$
\begin{equation}
q_\alpha \equiv k_\alpha L_\alpha / \pi.
\end{equation}
For a solution to exist, the determinant of ${\cal K}_\alpha$ should vanish, therefore the solutions are given by the null space of ${\cal K}_\alpha$. 
The solution space is spanned by eigenmodes corresponding to an infinite set $q_\alpha (i), i = 0, 1, 2, \ldots$ each  satisfying
\begin{equation}
\label{eq:bcfortextured}
\small
\frac{\pi}{\kappa_\alpha} ( 2 \cos^2 \frac{q_\alpha \pi}{2} - 1) + (1 - (\frac{q_\alpha \pi}{\kappa_\alpha})^2 - \sigma_\alpha^2) \cos \frac{q_\alpha \pi}{2} \sin \frac{q_\alpha \pi}{2} = 0.
\end{equation}
The smallest of  such $\{q_{\alpha} \}$, $q_{\alpha}(0)$ in each dimension contributes to the rate of the slowest mode as a function of sets of independent parameters $\kappa, \sigma, L$ in each dimension,
\begin{equation}
\small
\lambda_{a,0} (\{\kappa_\alpha\}, \{ \sigma_\alpha\}, \{ L_\alpha \} ) = \sum_\alpha \lambda_{a,\alpha}(0) \equiv \sum_\alpha  D (\frac{\pi}{L_\alpha})^2 q^2_{\alpha}(0)
\end{equation}
and takes the form of a set of parallel channels. 
Introducing the diffusion time 
\begin{equation}
\tau_\alpha^{-1} \equiv D (\frac{\pi}{L_\alpha})^2, 
\end{equation}
it can be put into the following form:
\begin{equation}
\label{eq:ratefortextured}
\lambda_{a,0} (\{\kappa_\alpha\}, \{ \sigma_\alpha\}, \{ \tau_\alpha \} ) = \sum_\alpha  \frac{q^2_{a,\alpha}(0)}{\tau_\alpha}.
\end{equation}

\subsection{{\it Solutions with a Symmetric Texture}}
For a symmetric texture ($\{ \sigma_\alpha = 0 \}$),  ${\cal K}_{\alpha}$ is
\begin{equation}
{\small 
\Big(   \begin{array}{cc}
      0 & \frac{\pi}{\kappa_\alpha} q_{\alpha} \cos \frac{q_{\alpha} \pi}{2} + \sin \frac{q_{\alpha} \pi}{2} \\
      \cos \frac{q_{\alpha} \pi}{2} - \frac{\pi}{\kappa_\alpha} q_{\alpha} \sin \frac{q_{\alpha} \pi}{2}  & 0 \\
   \end{array}
   \Big) }
\end{equation}
and the boundary condition factorizes.
For the slowest  rate, the symmetric solution should be taken 
\begin{equation}
\phi_{0} (\{ r_\alpha \} )  = \prod_{\alpha}^{d_e} A_{\alpha} \cos (k_{\alpha} r_\alpha)
\end{equation}
with the normalization condition
\begin{equation}
A_{\alpha}^2 \int_{-L_\alpha/2}^{L_\alpha/2} dr_\alpha \cos^2 (k_{\alpha} r_\alpha) = 1.
\end{equation}
The boundary condition yields
\begin{equation}
\label{eq:bcforsymmetric}
\cos  (q_{\alpha} \pi/2) - \frac{\pi}{\kappa_\alpha} q_{\alpha} \sin  (q_{\alpha} \pi/2)  = 0. 
\end{equation}
This leads to the class of solutions equivalent to those used by Brownstein and Tarr for the simple geometry. \citep{Brownstein:1979p779}
The rate for the slowest mode is then
\begin{equation}
\label{eq:rateforsymmetric}
\lambda_{s,0} (\{\kappa_\alpha\},\{ \tau_\alpha \} ) = \sum_\alpha  \frac{q^2_{\alpha}(0)}{\tau_\alpha}  .
\end{equation}
Note that it takes the form of three competing diffusion channels each of rate $\frac{1}{\tau_\alpha}$ weighted by $q_{s,\alpha}^2$ factor.
It is useful to further examine limiting behaviors in this symmetric case:
In the limit of $\kappa_\alpha \rightarrow 0,$
\begin{equation}
\label{eq:fastdiffusionlimit}
\lim_{\kappa_\alpha \rightarrow 0} q_{\alpha}(0) =  q^-_{\alpha} = \frac{\sqrt{2 \kappa_\alpha}}{\pi}
(1+ \frac{\kappa_\alpha}{4})^{-1/2}  + {\cal O}(\kappa_\alpha^{2})
\end{equation}
while in the opposite limit $\kappa_\alpha \rightarrow \infty$,
\begin{equation}
\label{eq:slowdiffusionlimit}
\lim_{\kappa_\alpha \rightarrow \infty} q_{\alpha}(0) =  q^+_{\alpha} =
1 -   \frac{2}{\kappa_\alpha}+ \frac{4}{\kappa_\alpha^2} + {\cal O}(\kappa_\alpha^{-3})
\end{equation}
where $\kappa_\alpha > 10$ is required for the expansion in $\frac{1}{\kappa_\alpha}$ to be accurate. 
In the special limit where all $\kappa_\alpha \ll 1,$
\begin{equation}
\label{eq:fastlimit}
\small
\lim_{\forall \kappa_\alpha \rightarrow 1}\lambda_{s,0} (\{\kappa_\alpha\},\{ \tau_\alpha \} ) 
\sim  \sum_\alpha  \frac{2 \bar\rho_\alpha}{L_\alpha}  (1 - \frac{\kappa_\alpha}{4}).
\end{equation}
Note that this approaches $(<\bar \rho> - <\bar \rho \, \kappa>/2)\frac{S}{V}$.
Under this condition, this form shows that the symmetric texture leads to that of the uniform case. 
In the opposite case ($\forall \kappa_\alpha \gg 1$), 
\begin{equation}
\label{eq:slowlimit}
\small
\lim_{\forall \kappa_\alpha \rightarrow \infty}\lambda_{s,0} (\{\kappa_\alpha\},\{ \tau_\alpha \} ) \sim  \sum_\alpha  \frac{1 - 2/\kappa_{\alpha}}{\tau_\alpha} .
\end{equation}
For a general symmetric case where either of the limits cannot be taken for all directions, i.e. when the $\kappa_\alpha$ parameters fall in the intermediate zone, appropriate interpolations may be made.

\begin{figure}[hpt] 
   \centering
   \includegraphics[width=3.in]{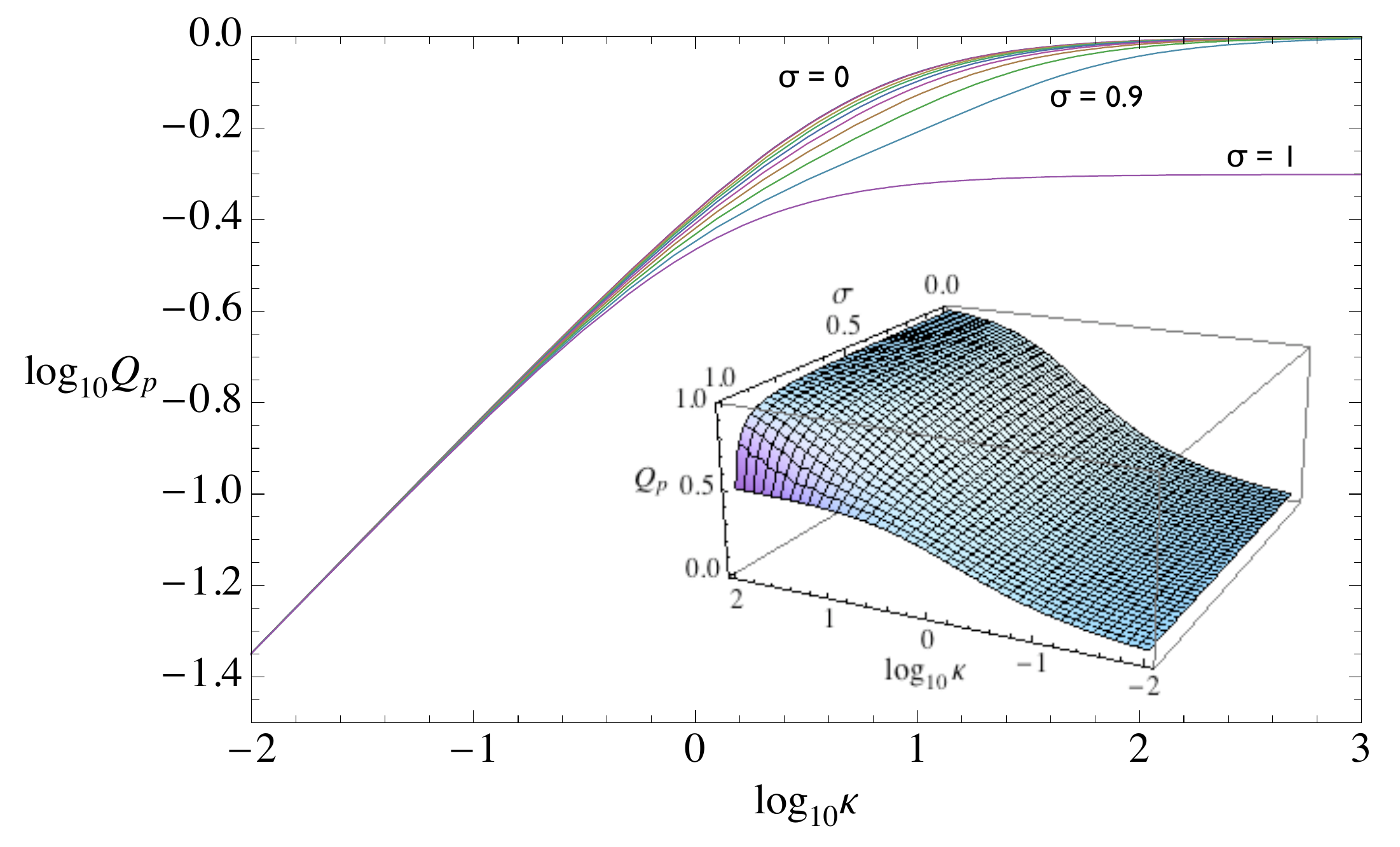} 
   \caption{\small $Q_p$ as a function of $\kappa$ and $\sigma$. The inset show a 3D plot of $Q_p(\kappa, \sigma)$. Note the rapid drop from $Q_p \sim 1$ to $0.5$ toward the large $\kappa$ and $\sigma \rightarrow 1$ corner, as it corresponds to doubling of the wavelength for the slowest mode and accompanying phase shift. Upper panel shows the series of curves $\log_{10}Q_p$ vs. $\log_{10}\kappa$ for values of $\sigma = 0., 0.1, 0.2, \ldots 1.0$. In the main panel,  all curves converge to a $Q_p \sim \kappa$ behavior for $\kappa < 0.1$.}
   \label{fig:Q3D}
\end{figure}

\subsection{{\it Cube with Uniform $\rho$}}
The solutions obtained above for general $\epsilon_\alpha$-$\sigma_\alpha-L_\alpha$ variations, may be compared against that of a pore with globally uniform $\rho = \rho_g$ and cubic geometry with the size, $L_g^{d_e}$ that is equal to that of the rectangle $\prod_\alpha^{d_e} L_\alpha$.  
Defining the effective $\kappa_g$ 
\begin{equation}
\label{eq:definekappag}
\kappa_g \equiv <\bar \rho> \frac{L_g}{D},
\end{equation}
it is related to $\kappa_\alpha$'s  via 
\begin{equation}
\kappa_g  = 2 \sum_\alpha  \frac{\kappa_\alpha}{L^2_\alpha}  \frac{V L_g}{S}
\end{equation}
using the fact that $S_\alpha L_\alpha = V$ for any $\alpha$ for a rectangle.
The slowest rate for the globally uniform cubic pore takes the simple form: 
\begin{equation}
\label{eq:rateforglobal}
\lambda_{g,0} (\kappa_g) = d_e  \frac{q^2_{g}(0)}{\tau_g} 
\end{equation}
where $q_{g}(0)$ is the smallest of the $q_{g}$'s that satisfy:
\begin{equation}
\label{eq:bcforglobal}
\cos  (q_{g} \pi/2) - \frac{\pi}{\kappa_g} q_{g} \sin  (q_{g} \pi/2)  = 0
\end{equation}
and we introduced 
\begin{equation}
\tau_g^{-1} = D (\frac{\pi}{L_g})^2.
\end{equation}
Instead of the volume matching criterion, had we chosen $L_g = 2 d_e (S/V)^{-1}$ so that the cube has the same surface-to-volume ratio to that of the rectangle, we recover the expected behavior that $\lambda_{g,0}(\kappa_g) \rightarrow \lambda_{0}(\kappa_\alpha,0)$ of the symmetric case when $\kappa_\alpha, \kappa_g  \ll 1$  and $\bar \rho_\alpha = <\bar \rho>$.  
However, in many contexts, it makes more sense to impose the equivalent-volume criterion, i.e. deformation of pore geometry while preserving overall porosity. 
In the following, we will employ such a criterion with $L_g^{d_e} = V$.

\subsection{{\it Recipe for Estimating Bounds}}
In summary, we note the following: 
For a cube of side length $L_g$ with a uniform $\rho_g$, the rate, $\lambda_{g,0}$, is controlled by the one-dimensional condition (Eq.\ref{eq:bcforglobal}) weighted by the arithmetic mean of the geometrical factors i.e. diffusion rates (Eq.\ref{eq:rateforglobal}). In the symmetric case (rectangle with $\epsilon_\alpha$'s), the boundary conditions become distinct for each direction (Eq.\ref{eq:bcforsymmetric}), controlled by the $\kappa_\alpha$ factor; the slowest rate, $\lambda_{s,0}$, is given by 
Eq.\ref{eq:rateforsymmetric} and is a function of $\kappa_\alpha$ and $\tau_\alpha$'s. The uniform $\rho_g$ is related to the $\bar\rho_\alpha$'s through Eq.\ref{eq:globalrho}. 
In the asymmetric case ($\sigma_\alpha \ne 0$ for at least one dimension), the even and odd modes mix (i.e. phase shifts) with its degree controlled by $\sigma_\alpha$, Eq.\ref{eq:bcfortextured}.  The resulting rate for the slowest  decay mode (Eq.\ref{eq:ratefortextured}), $\lambda_{a,0}$, is a function of $\kappa_\alpha, \tau_\alpha$ as well as $\sigma_\alpha$'s.
The diffusive rates $\tau^{-1}_\alpha$'s control the purely geometrical aspects of the pore, $L_\alpha$.

\begin{figure}[h] 
   \centering
   \includegraphics[width=3.in]{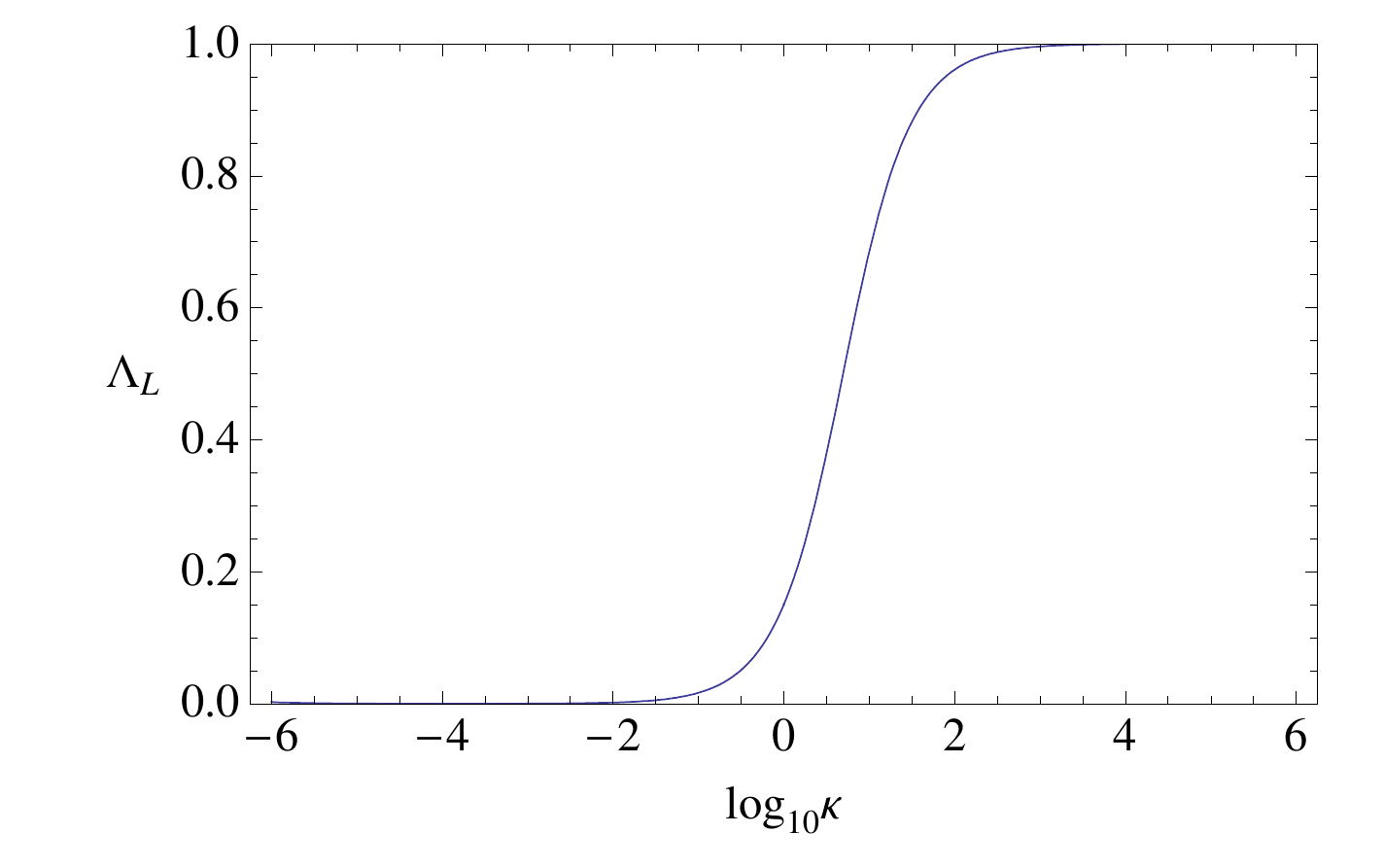} 
   \caption{\small $\Lambda_L$ as a function of $\kappa$. It is vanishingly small for $\kappa < 0.01$ but increases toward the value of $1$ for large $\kappa$.  }
   \label{fig:LambdaL}
\end{figure}
If one is interested in the change of the rate in going from the base (uniform, isotropic) cube to the  rectangular pore with a symmetric texture,  its fractional shift, $\delta \lambda_{g:s,0} \equiv \lambda_{g,0} - \lambda_{s,0}$, is given by 
\begin{equation}
\label{eq:delta1}
\triangle_{s} \equiv\frac{\delta \lambda_{g:s,0}}{\lambda_{g,0}} =  \frac{1}{d_e} 
\sum_\alpha  ( 1 - \frac{\tau_g}{\tau_\alpha}  \frac{q^2_{s,\alpha}(0)}{q^2_{g}(0) })
\end{equation}
where the subscripted $q_{s,\alpha}$ is used to indicate that it is the solution of the symmetric case.
For comparison between the asymmetric and the reference cases with $\delta \lambda_{g:a,0} \equiv \lambda_{g,0} - \lambda_{a,0}$, we obtain
\begin{equation}
\label{eq:delta2}
\triangle_{a} \equiv\frac{\delta \lambda_{g:a,0}}{\lambda_{g,0}} =  \frac{1}{d_e} 
\sum_\alpha  ( 1 - \frac{\tau_g}{\tau_\alpha}  \frac{q^2_{a,\alpha}(0)}{q^2_{g}(0) }).
\end{equation}
These expressions treat the impact of both geometrical and $\rho$-textural changes on an equal footing. Separation of their individual impact is straightforward: For geometrical impact alone, one may take the first expression evaluated with $\epsilon_\alpha = 0$ for all $\alpha$ for a given rectangular geometry (let us identify this special case with subscript $\tilde{s}$ with  $\bar\rho_\alpha = \rho_g$ for all $\alpha$, and $\kappa_\alpha = \rho_g L_\alpha / D$). Comparing this to the cubic, uniform $\rho$ case, 
\begin{equation}
\label{eq:delta3}
 \triangle_{geom} \equiv\frac{\delta \lambda_{g:\tilde{s},0}}{\lambda_{g,0}} =\frac{1}{d_e} 
\sum_\alpha  ( 1 - \frac{\tau_g}{\tau_\alpha}  \frac{q^2_{\tilde{s},\alpha}(0)}{q^2_{g}(0) }).
\end{equation}
For the difference between the $\rho$-textured and a uniform $\rho$ with the same rectangular geometry, 
use the $\tilde{s}-$case as the reference instead; the fractional shift between the uniform and textured rectangle is
\begin{equation}
\label{eq:delta4}
 \triangle_\rho \equiv \frac{\delta \lambda_{\tilde{s}:a,0}}{\lambda_{\tilde{s},0}} =  \frac{\sum_\alpha  \frac{q^2_{\tilde{s},\alpha}(0)}{\tau_\alpha} (1 -  \frac{q^2_{a,\alpha}(0)}{q^2_{\tilde{s},\alpha}(0)} ) }{\sum_\alpha  \frac{q^2_{\tilde{s},\alpha}(0)}{\tau_\alpha}}.
\end{equation}
Due to the weighting factor $\lambda_{\tilde{s},\alpha}(0)/\sum_\alpha \lambda_{\tilde{s},\alpha}(0)$, the largest $\lambda_{\tilde{s},\alpha}(0)$ will dominate, which in the $\tilde{s}-$ system, is equivalent to the shortest $L_\alpha$ since $\bar\rho_\alpha = <\bar\rho>$ for all $\alpha$. Therefore, for an extremely anisotropic geometry, $L_1 \ll L_\alpha$ $(\alpha = 2, \ldots d_e)$, such as a slab-like pore, this reduces to 
$(1 - q_{a,1}^2(0) / q_{{\tilde s},1}^2 (0))$, dominated by the $\rho-$variation in the most constricted dimension.  Note that while
$\delta\lambda_{g:a,0}= \delta\lambda_{g:\tilde{s},0} +\delta\lambda_{\tilde{s}:a,0}$ is true, 
one cannot erroneously assume 
$\triangle_a =  \triangle_{geom} +  \triangle_{\rho}$.

For a general variation of $\{\bar \rho_\alpha, \sigma_\alpha, L_\alpha \} (\alpha = 1, \ldots d_e)$ values,  we have a $3\times d_e$ dimensional phase space. 
However, one can determine the impact of moving in such a space by reading a few numbers off a universal function $Q_p(\sigma, \kappa)$. 
First, pure geometrical aspects are incorporated in terms of $\tau_\alpha$'s once the aspect ratio of the pore is set.
Next, more subtle aspect involving both the geometry and the $\rho$ texture is addressed via 
the series of master curves $Q_p(\kappa,\sigma)$ from the smallest $|Q|$ value that satisfies the condition 
\begin{equation}
\label{eq:defineQp}
\small
\frac{\pi}{\kappa} ( 2 \cos^2 \frac{Q \pi}{2} - 1) + (1 - (\frac{Q \pi}{\kappa})^2 - \sigma^2) \cos \frac{Q \pi}{2} \sin \frac{Q \pi}{2} = 0.
\end{equation}
This  establishes a manifold in the $\kappa-\sigma-Q$ space, plotted in  Figure \ref{fig:Q3D}, that determines the slowest rate for any configuration prescribed in Figure \ref{fig:rectangles}. Obviously, one can further construct manifolds that correspond to faster modes, although logistics of tracking among closely spaced higher eigenvalues may not be trivial in practice. Both $q_g(0)$ and $q_{s,\alpha}(0)$ of Eq.\ref{eq:delta1} can be read off from such $Q_p$ at respective values of $\kappa = \kappa_g$(Eq.\ref{eq:definekappag}) and $=\kappa_\alpha$(Eq.\ref{eq:definekappaalpha}) with $\sigma = 0$. 
Likewise, the values of $q_{a,\alpha}(0)$ are given from $Q_p$ at $\kappa= \kappa_\alpha$(Eq.\ref{eq:definekappaalpha}) and $\sigma = \sigma_\alpha$(Eq.\ref{eq:definesigmaalpha}). 

\begin{figure}[h] 
   \centering
   \includegraphics[width=2.7in]{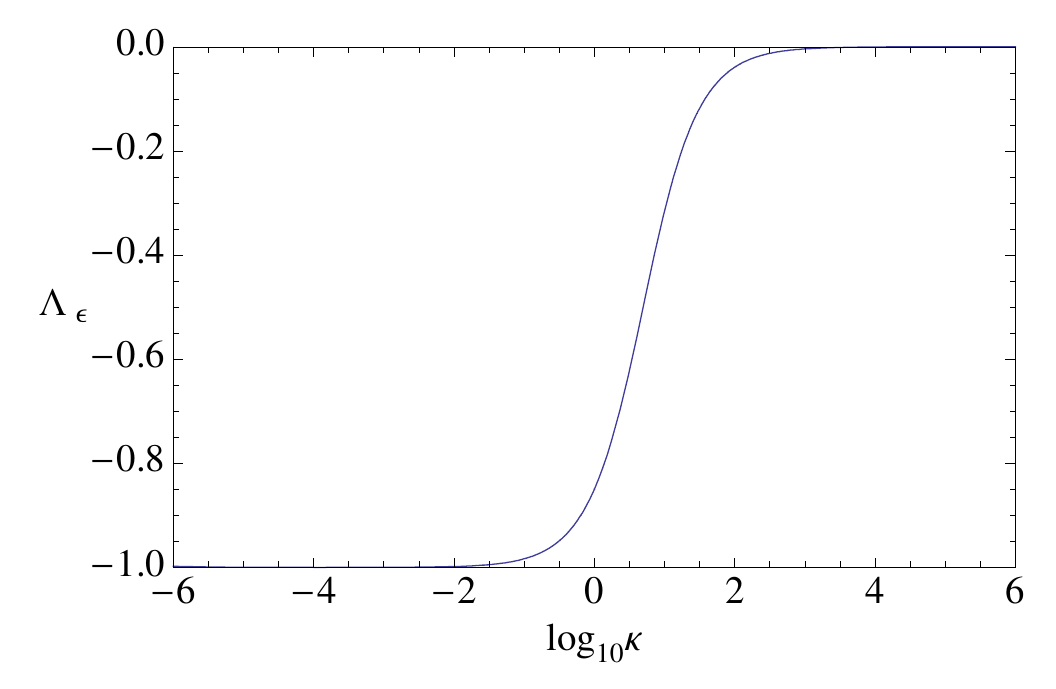} 
   \includegraphics[width=2.9in]{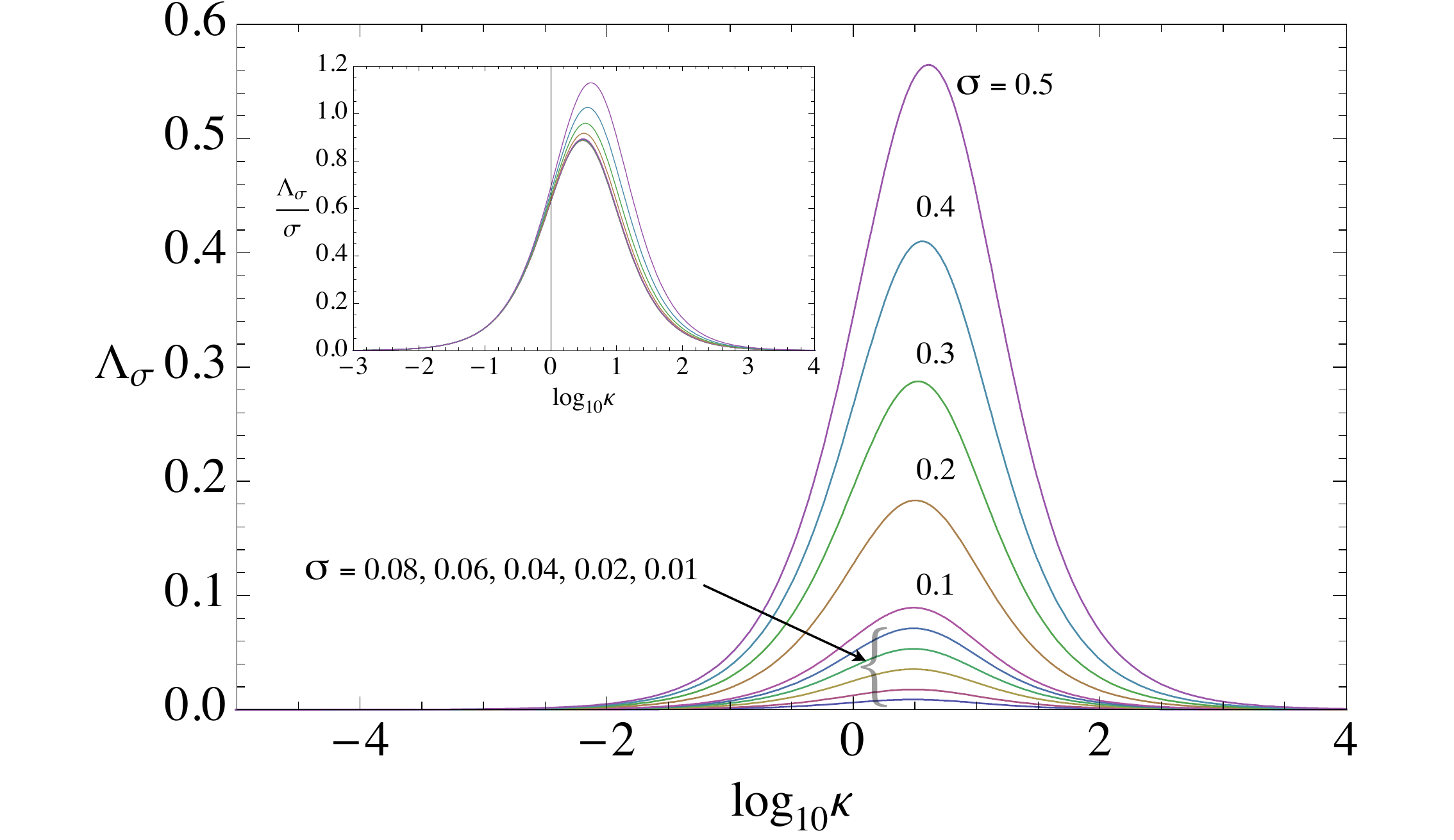} 
   \caption{\small \small
   Top: $\Lambda_\epsilon$ as a function of $\kappa$. It is non-negligible for small $\kappa$ values ($\sim - 1$) and becomes negligible for large $\kappa > 100$. Even for small $\kappa$, however,  the overall contribution from $\Lambda_\epsilon$ averages out due to the condition $\sum_\alpha \frac{\epsilon_\alpha}{L_\alpha} = 0$. Note that the functional form is also essentially identical to $\Lambda_L$, reflecting the fact that the symmetric $\rho$ texture (i.e. $\epsilon_\alpha$) has an effect similar to that from geometrical dilation/contraction of the pore dimension. Bottom: $\Lambda_\sigma$ as a function of $\kappa$ and for values of $\sigma = 0.01., 0.02, 0.04, 0.06, 0.08, 0.1, 0.2, 0.3, 0.4, 0.5$. The inset shows $\Lambda_\sigma/\sigma$. For $\sigma \le 0.1$, the curves tend to converge toward the universal form for the whole range of $\kappa.$}
   \label{fig:LambdaEpsilonAndSigma}
\end{figure}

As an application of this recipe, let us consider three hypothetical variations: one in geometry ($\tau_\alpha$) alone and the others in terms of $\epsilon_\alpha$ and $\sigma_\alpha$'s in comparable fractions, say $50 \%$ to get an idea of their relative significance. (I) For a pure {\em pancake-like} geometrical deformation specified by $\tau_x \rightarrow 2 \tau_g$ and $\tau_y, \tau_z\rightarrow \tau_g/ \sqrt{2}$, the rate becomes faster with 
$\triangle_{geom} = -0.222$  $(\kappa_g = 0.32)$, $-0.286 $ $(\kappa_g = 1.6)$, 
$-0.435$  $(\kappa_g = 6.4)$. Now, consider a cubic pore with variations in $\rho$ only: (II) For a symmetric $\delta \rho$ texture with $\epsilon_x = -0.5$, $\epsilon_y, \epsilon_z = 0.25$, the rate slows down with $\triangle_{s} = 0.0062  (\kappa_g = 0.32)$, $0.024  (\kappa_g = 1.6)$ and $0.039  (\kappa_g = 6.4)$. These small values are due to cancellation of larger numbers from each direction (e.g. with $\kappa = 6.4,$ we had $\triangle_{s,x} = 0.1016$ and $\triangle_{s,y} , \triangle_{s,z}  = -0.0311$). (III) For an asymmetric texture with $\sigma_\alpha = 0.5$ for all $\alpha$'s, it always slows down with  $\triangle_{a} = 0.034 (\kappa_g = 0.32)$, $0.099 (\kappa_g = 1.6)$, $0.106 (\kappa_g = 6.4)$.

\subsection{{\it Small $\kappa$ limit}}
Without going into specific excursions in the $3\times d_e$-dimensional phase space, we can make general observations on how changes in pore aspect ratio and $\rho$ textures affect the rates. 
Consider a rectangle of $\{L_\alpha\}$ ($S/V = \sum_\alpha 2/ L_\alpha$) with an asymmetric textured $\{ \bar \rho_\alpha \}$ and $\{\sigma_\alpha\}$ values. Consider also its variant, the symmetric system in which $\forall \sigma_\alpha = 0$.
We also construct an equivalent cube of side $L_g$ and globally uniform $<\bar \rho>$. 
From Eq.\ref{eq:rateforglobal}, the rate for the globally uniform cube is determined by $Q_p^2(\kappa_g, 0)$.
Introduce aspect ratio variations, then the fractional difference in the rate (Eq.\ref{eq:delta1}) between the symmetric and the cube is controlled by  the $(1 - \frac{\tau_g}{\tau_\alpha} Q_p^2(\kappa_\alpha,0)/Q_p^2 (\kappa_g,0))$ factors. In the limit where all $\kappa$'s are small (Eq.\ref{eq:fastdiffusionlimit}), they become $(1 - (\frac{L_g}{L_\alpha})^2 \frac{\kappa_\alpha}{\kappa_g}\frac{1+ \kappa_g/4}{1+ \kappa_\alpha/4})$, and using Eq.\ref{eq:definekappag}, we can show 
\begin{equation}
\label{eq:trigrfastdiffusionlimit}
\lim_{\forall \kappa_\alpha \rightarrow 0}\triangle_{s} \sim 1-  \frac{1}{ d_e } \sum_\alpha  \frac{L_g}{L_\alpha} (1 +  \epsilon_\alpha ) + {\cal O}(\kappa_g) .
\end{equation}
Furthermore, due to the property Eq.\ref{eq:conditionforepsilon}, the term linear in $\epsilon_\alpha$ vanishes.
Therefore, we have 
\begin{equation}
\label{eq:trigrfastdiffusionlimit2}
\lim_{\forall \kappa_\alpha \rightarrow 0}\triangle_{s} \sim 1-  \frac{1}{ d_e } \sum_\alpha  \frac{L_g}{L_\alpha} = \frac{\delta S}{S} + {\cal O}(\kappa_g) .
\end{equation}
where the last equality follows from  the porosity preserving definition $L_g = V^{1/d_e}$. 
The expression above (and more generally Eq.\ref{eq:delta1}) reveals the separation of the geometry (through $\frac{1}{L_\alpha}$ factors) and the class of $\delta \rho(\r)$ variation (via vanishing of the $\epsilon_\alpha / L_\alpha$ factors), at least in the small $\kappa_g$ limit under the fixed volume condition. Note that with $\epsilon_\alpha$-type variations, the slowest modes are still all symmetric functions in each dimension. Volume preservation requires that both compression ($\epsilon_\alpha >0$) and dialation($\epsilon_\alpha <0$) should be present in different directions, and as a result, their impact cancels out to first order. Even if it survives, it is eclipsed by the impact of  change in the surface-to-volume ratio. 
For large values of $\kappa_g$, terms involving $\epsilon_\alpha$'s may survive to contribute, and $\triangle_{s}$ should be numerically evaluated out of the master curve $Q_p^2(\kappa,0)$ evaluated at $\kappa = \kappa_g$ and $\kappa_\alpha$'s, each multiplied by the geometrical factor, $(\frac{L_g}{L_\alpha})^2$. 

\begin{figure}[h] 
   \centering
   \includegraphics[width=2.8in]{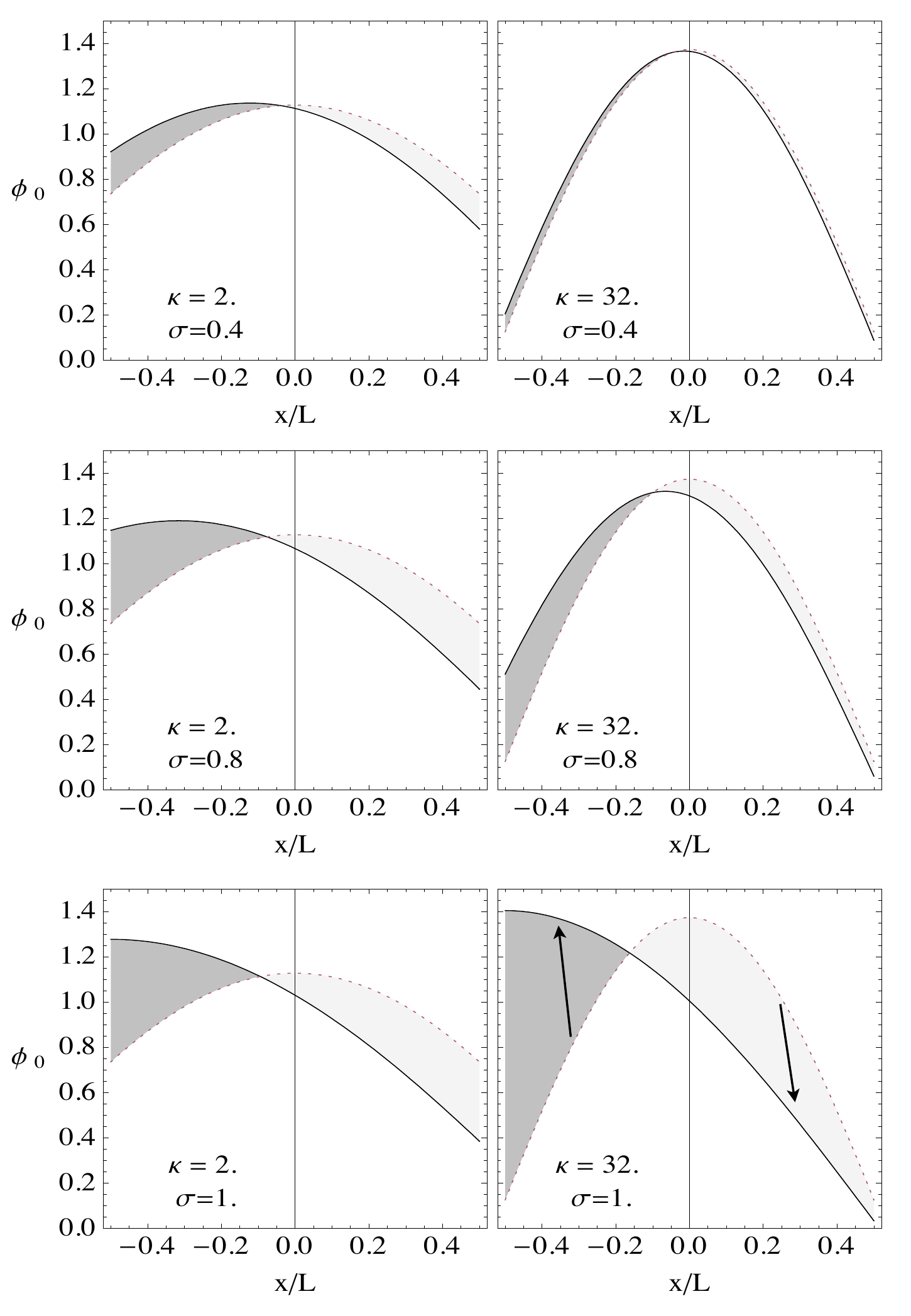} 
   \caption{\small Changes in the profile of the slowest eigenmode $\phi_0 (\r)$ for different values of $\kappa$ and $\sigma$.  We assume all $\epsilon_\alpha$ and $\sigma_\alpha$'s are zero except in the x-direction for which $\sigma_x \ne 0$ as indicated for the solid curves. The broken curves are for the uniform case with $\sigma_x = 0$. The shades indicate whether the profile increases (darker) or decreases (lighter) when $\sigma_x \ne 0$. }
   \label{fig:ProfileForVariousSigmaAndKappas}
\end{figure}

For the asymmetric texture ($\sigma_\alpha \ne 0$),  the mode profile, in addition to the contraction/dialation,  goes through a phase-shift in general. ($B_\alpha \ne 0$ in Eq.\ref{eq:groundstate}) as it is being pulled at one end and pushed at the other by the asymmetric $\rho_{\alpha\pm}$. (See also Figure \ref{fig:ProfileForVariousSigmaAndKappas} and discussion below) The fractional change in the rate is constructed from reading off $Q_p^2$ curves, and it involves moving across varying $\sigma_\alpha$'s, i.e. excursions in the vertical direction in Figure \ref{fig:Q3D}.

\begin{figure}[h] 
   \centering
   \includegraphics[width=3.in]{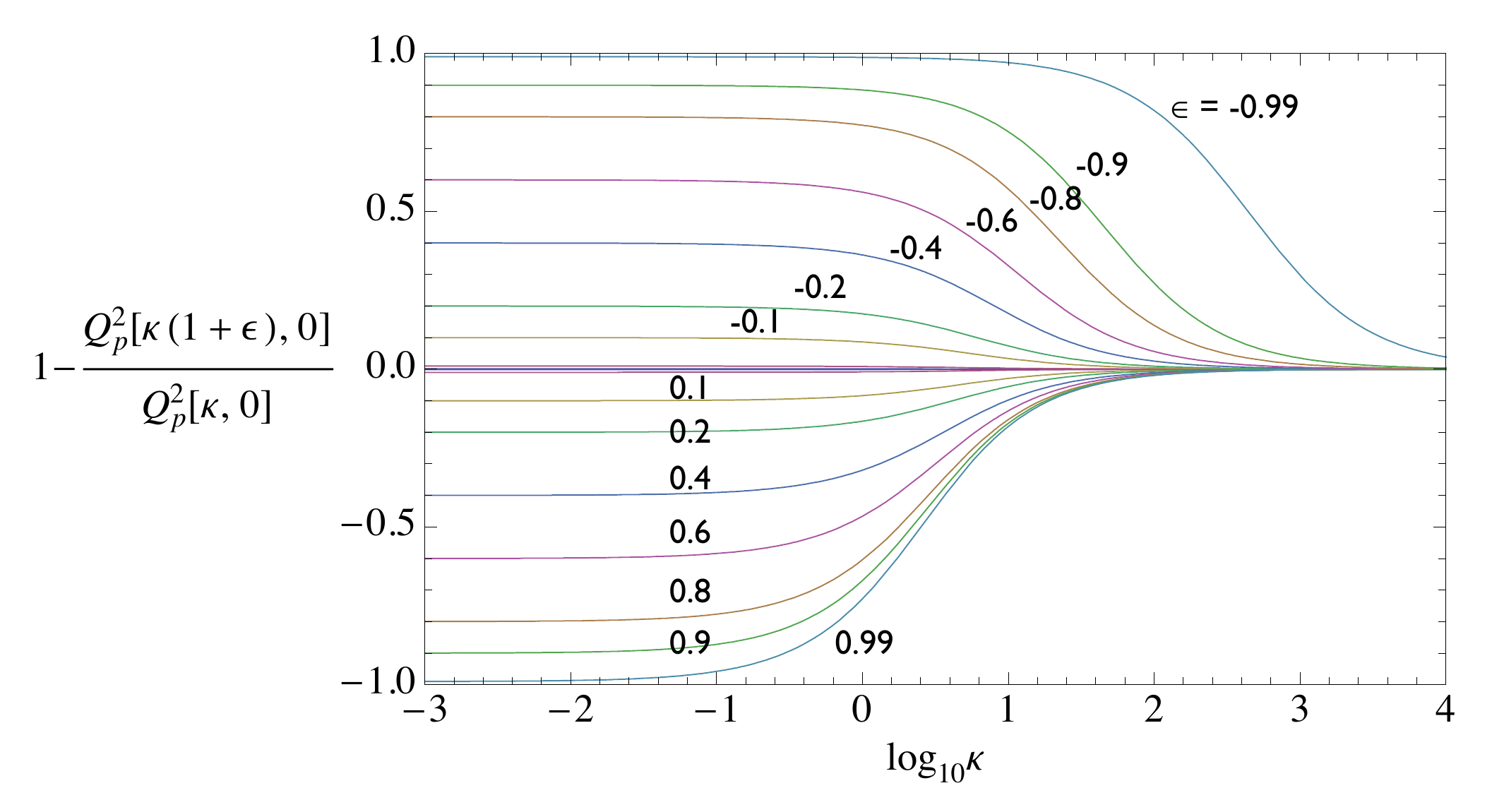} 
   \includegraphics[width=3.in]{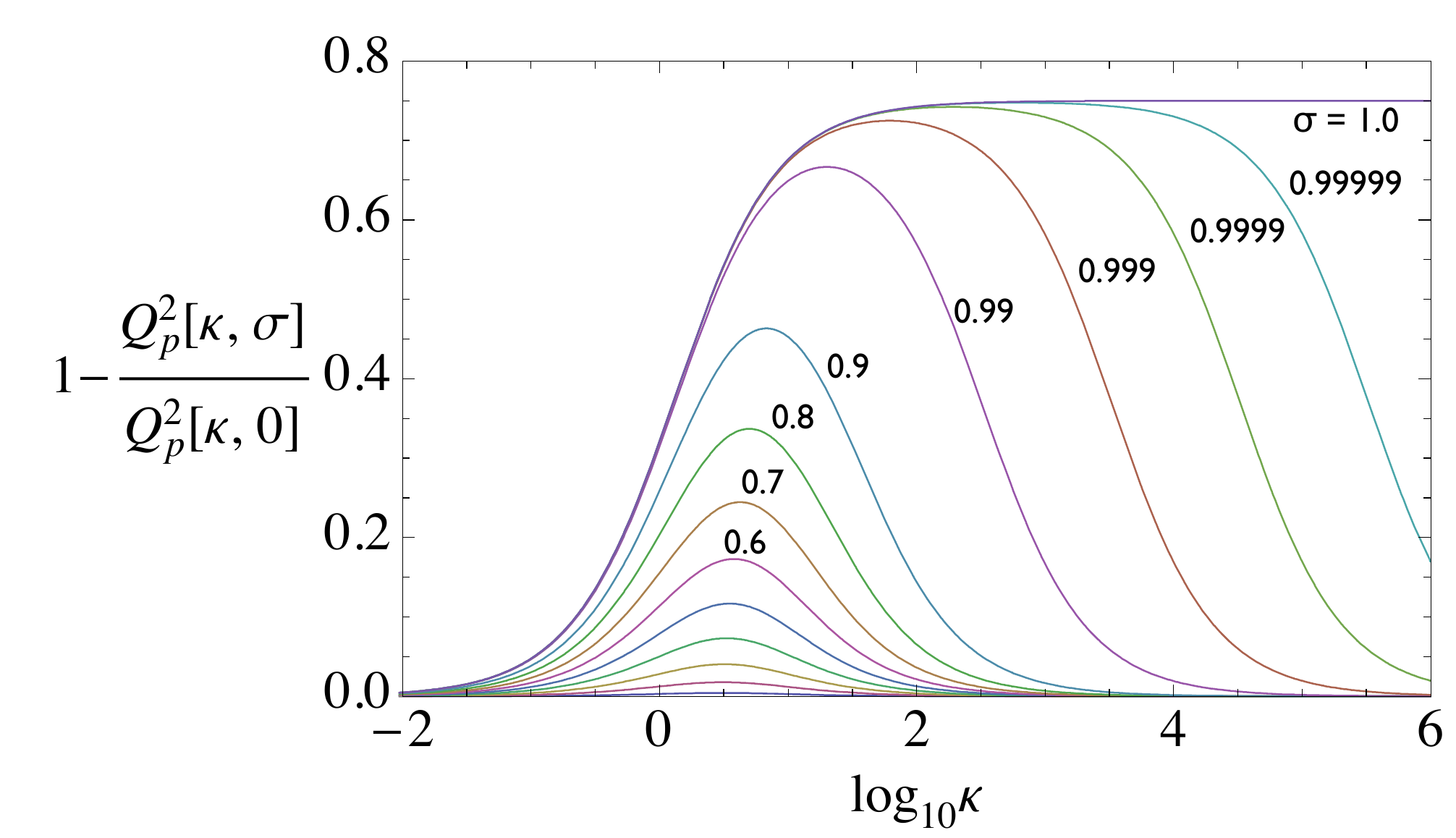} 
   \caption{\small Top: Fractional shift factor for the symmetric texture with $\sigma_\alpha = 0$ and $-1 < \epsilon <1$. By definition of $<\bar \rho >$ and $\epsilon_\alpha$, we have $\epsilon = 0$ for 1-dimensional system, and for $d_e \ge 2$, $\sum_\alpha (\epsilon_\alpha  / L_\alpha -  \delta L_\alpha / L_\alpha^2) = 0$ should be satisfied for a general symmetric case. 
   Bottom: Fractional shift factor for the texture with $1 \ge \sigma_\alpha > 0$, $\epsilon_\alpha = 0$. No assumptions are made about the size of $\sigma$. }
   \label{fig:LambdaEpsilonSigmaLarge}
\end{figure}

\begin{figure}[h] 
   \centering
   \includegraphics[width=3in]{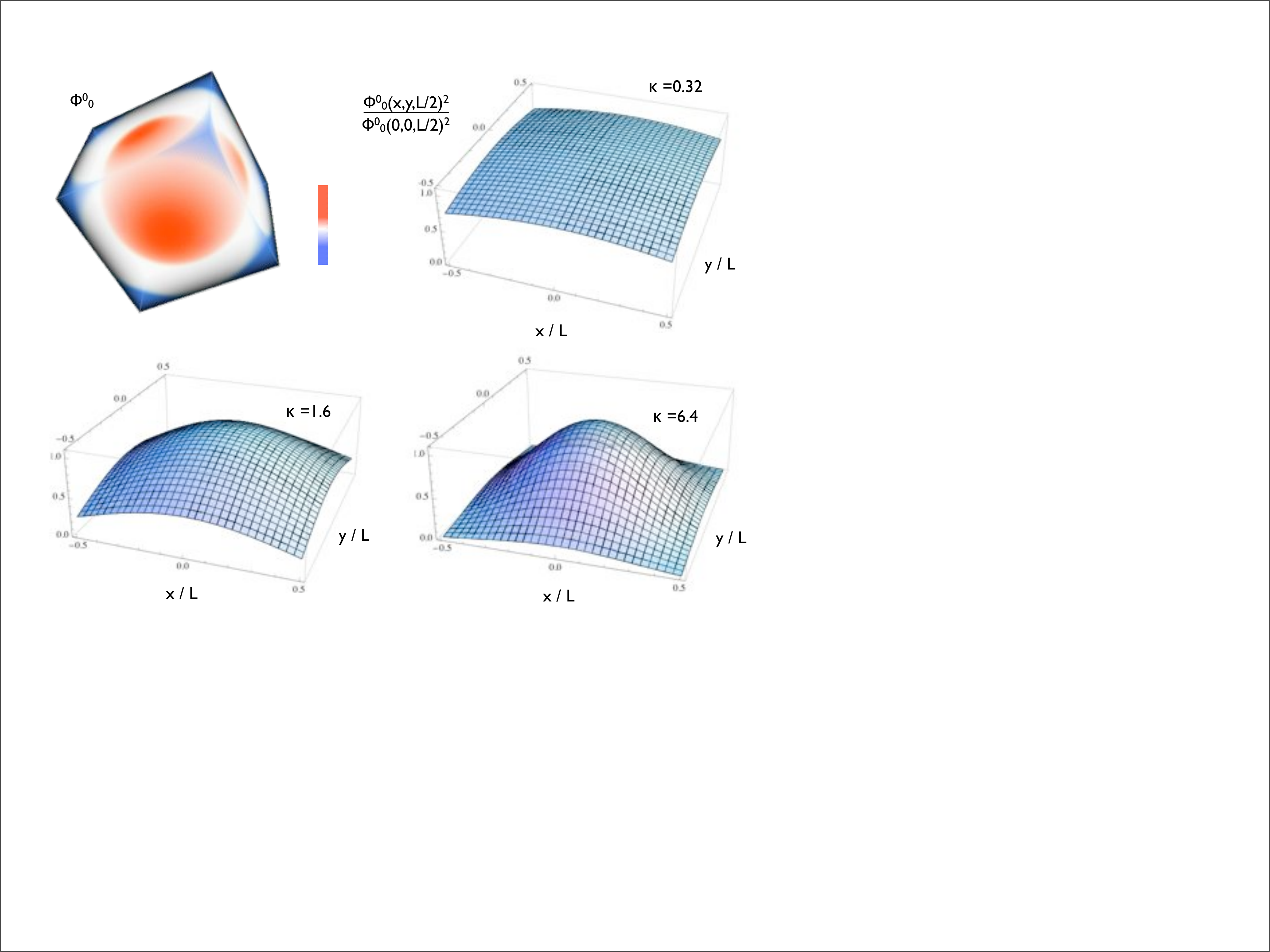} 
   \caption{\small Top-left panel: Profile of $\phi_0^0(\r)$ for $\kappa = 1.6$ for a cubic pore. The colormap was chosen to emphasize the depletion of $\phi_0^0$ near the eight corners of the cube. The rest shows the profile of $(\phi_0^0)^2$ in one of the interfacial planes normalized its value at the center of the plane. $\kappa = 0.32, 1.6$ and $6.4$.
 }
   \label{fig:groundstate}
\end{figure}

\begin{figure}[h] 
   \centering
   \includegraphics[width=3in]{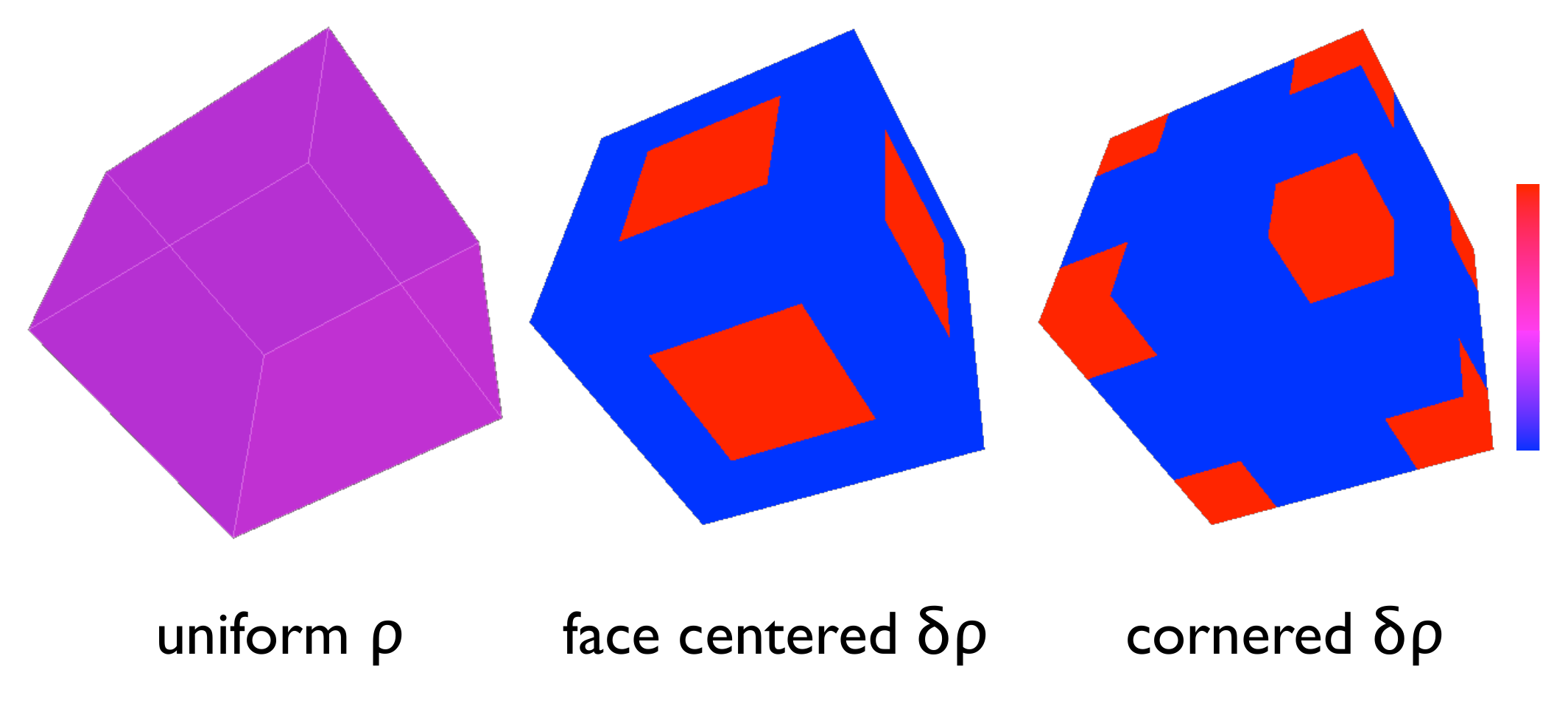} 
   \caption{\small Three $\rho$ textures considered for comparison. The first panel shows a cube with a uniform $\rho_0$.  Two non-uniform $\rho$ textures are shown. The mid-panel shows the {\em face-centered} $\delta \rho$ which has an enhanced $\rho$ in the center of each face. The right panel shows the {\em cornered} $\delta \rho$ which has enhanced $\rho$ in the eight corners of the cube. The total area of the red is chosen to be equal in both cases.}
   \label{fig:drhoCube}
\end{figure}

\begin{figure}[h] 
   \centering
   \includegraphics[width=2.5in, height=1.5in]{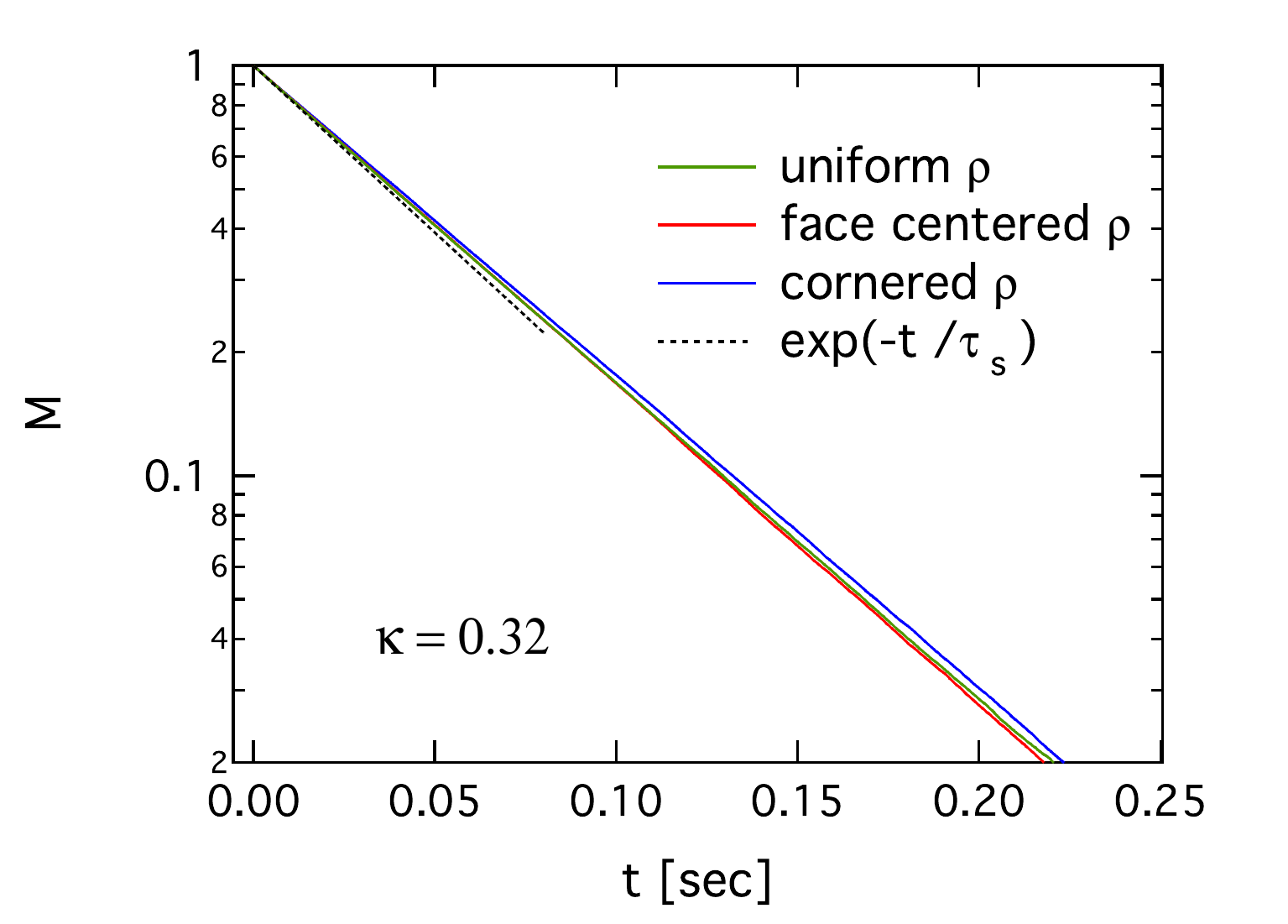} 
  \includegraphics[width=2.5in, height=1.5in]{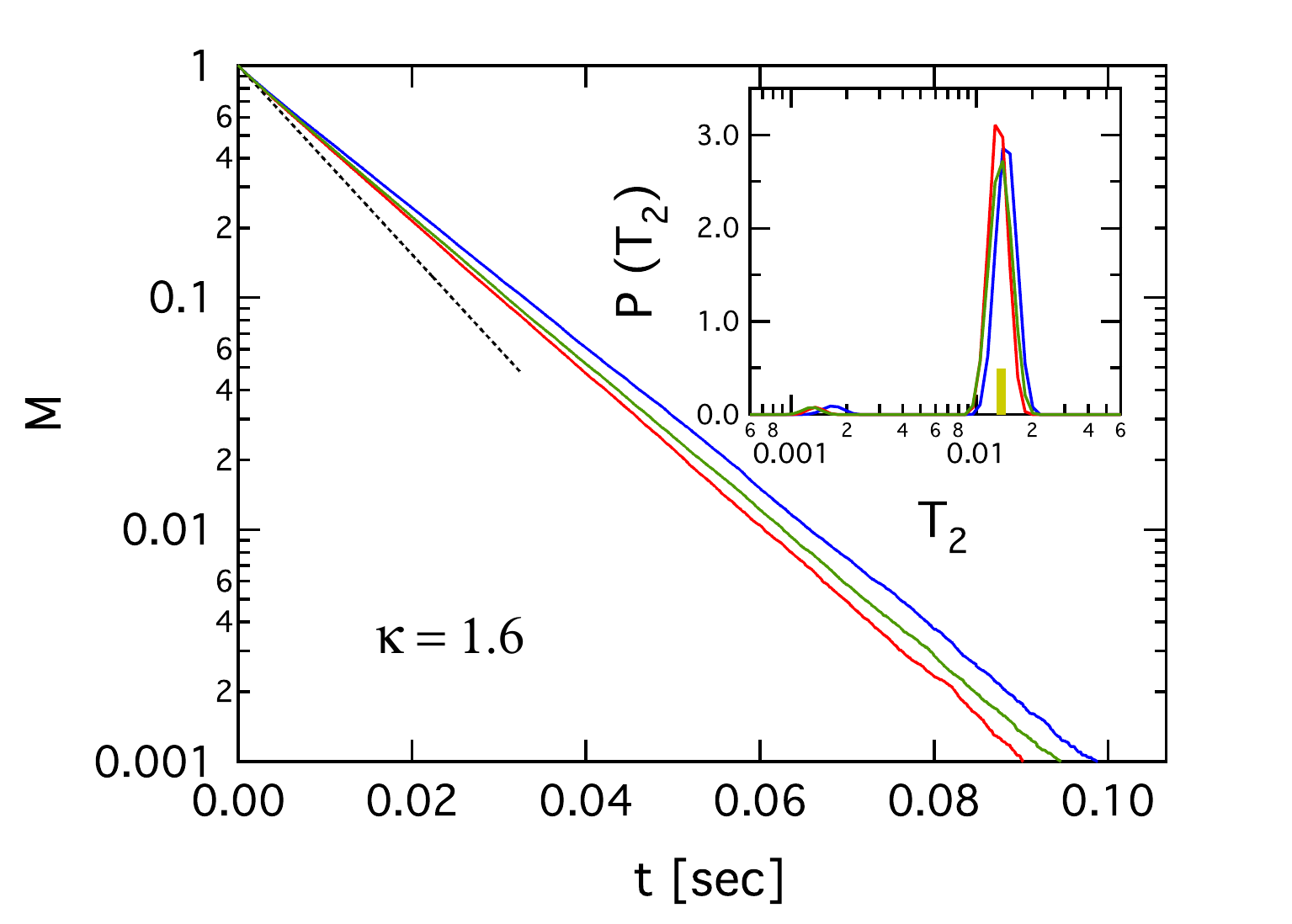} 
  \includegraphics[width=2.5in, height=1.5in]{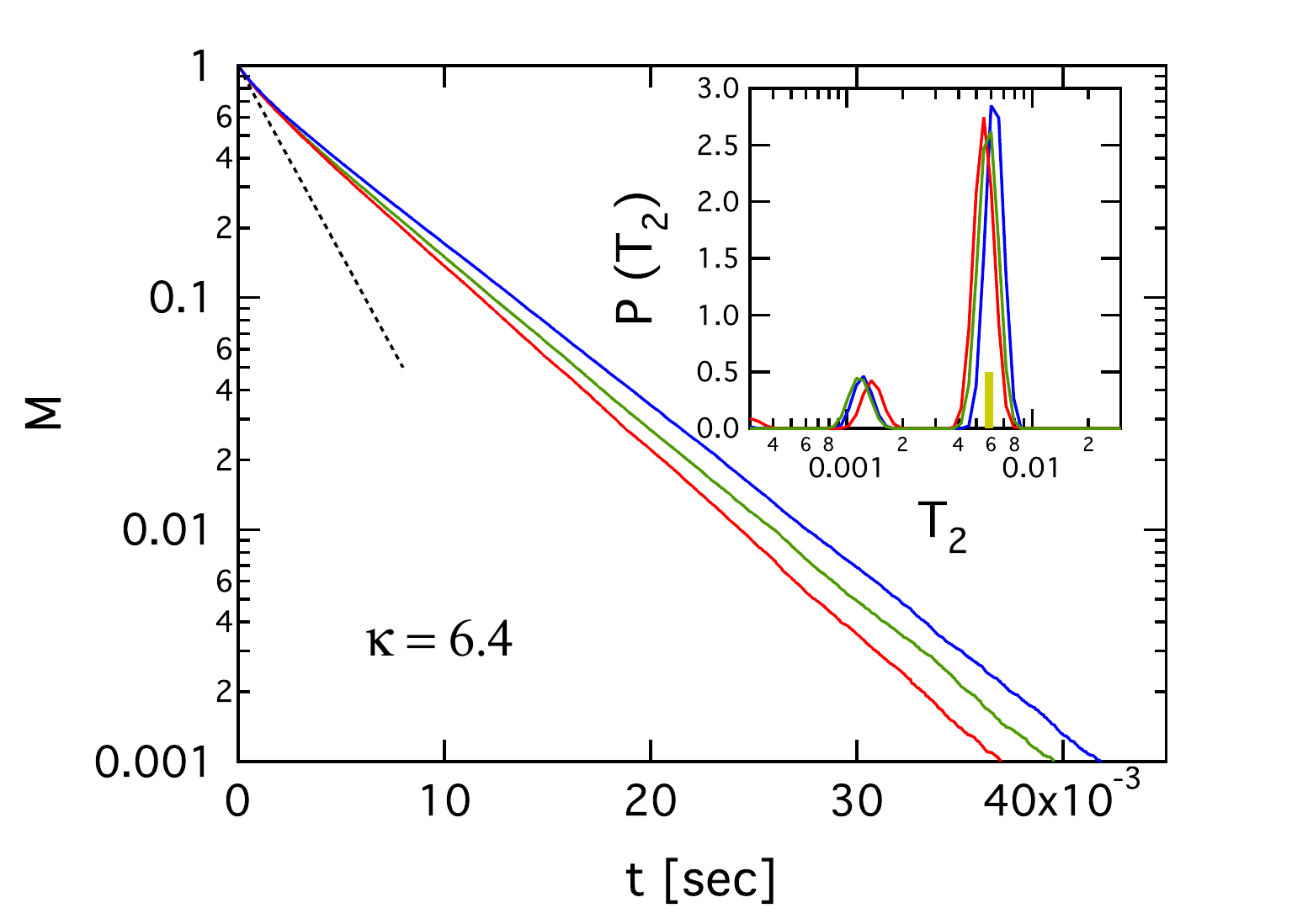} 
   \caption{\small Comparison of NMR responses simulated for the three $\rho$ textures of Figure \ref{fig:drhoCube}. The first three panels (each with $\kappa = 0.32, 1.6, 6.4$) show $M$ vs. $t$ with uniform $\rho$ (green), face-centered $\delta \rho$ (red) and cornered $\delta \rho$ (blue). The broken lines for short t indicate the predicted initial slope of the decay curve, $\sim e^{-t / \tau_s}$ for each $\kappa$. The insets for $\kappa = 1.6, 6.4$ show the Laplace-inversion ($T_2$-distribution) results with the same color scheme. Also shown in the insets are the positions of the slowest $T_2$ from exact calculation indicated by the short vertical bar in yellow.}
   \label{fig:drhoCubeSimResults}
\end{figure}

\subsection{{\it Large $\kappa$ limit}}
\label{sec:largekappa}
Let us now examine geometrical and textural deformations off an arbitrary value of $\kappa_g$ which is {\em not necessarily} assumed small.
For small fractional changes in $\kappa_\alpha$'s arising from a geometrical deformation alone, Eq.\ref{eq:delta3} yields
\begin{equation}
\label{eq:smallgeometricaldeformation}
 \triangle_{geom}  =   \frac{1}{d_e} \sum_\alpha \frac{ d L_\alpha }{L_\alpha}(1 + \Lambda_L (\kappa_g)).
\end{equation}
where 
\begin{equation}
\label{eq:defineLambdaL}
\Lambda_L (\kappa) =  1 -  2\kappa
\frac{d Q_p (\kappa,0)/d\kappa}{Q_p(\kappa, 0)}
\end{equation}
which is plotted in Figure \ref{fig:LambdaL}. In the small $\kappa$ limit, it converges to the value of $0$, making $ \triangle_{geom} $ converge to $\delta S / S$ as expected, while in the opposite limit, the convergence is toward  the value of $1$, so that it gradually accommodates the {\em slow diffusion} asymptote $\lambda_\alpha(0) \propto \frac{1}{L_\alpha^2}$.

In the case of changes arising from $\rho$ texture alone, we take the $\tilde{s}-$ state as our reference with $q_{\tilde{s},\alpha}$'s solely determined from $\rho_g$ and $\{ L_\alpha \}$, and with $\epsilon_\alpha \ll 1, \sigma_\alpha \ll 1$,
\begin{equation}
\label{eq:smalltexturaldeformation}
\triangle_{\rho} \rightarrow  \frac{\sum_\alpha  \frac{q^2_{\tilde{s},\alpha}(0)}{\tau_\alpha} 
( \epsilon_\alpha  \Lambda_\epsilon (\kappa_\alpha) +  \sigma_\alpha \Lambda_\sigma (\kappa_\alpha , \sigma_\alpha) )  }{\sum_\alpha  \frac{q^2_{\tilde{s},\alpha}(0)}{\tau_\alpha}}
\end{equation}
where we define
\begin{equation}
\label{eq:defineLambdaEpsilon}
\Lambda_\epsilon (\kappa) = - 2 \kappa \,  \partial \ln Q_p (\kappa, 0) / \partial \kappa 
\end{equation}
and 
\begin{equation}
\label{eq:defineLambdaSigma}
\Lambda_\sigma (\kappa, \sigma) = - 2  \partial \ln Q_p (\kappa , \sigma) / \partial \sigma .
\end{equation}
Figure \ref{fig:LambdaEpsilonAndSigma} show the functions $\Lambda_\epsilon(\kappa)$ and $\Lambda_\sigma(\kappa, \sigma)$ respectively that can be used for small values of $\epsilon_\alpha$ and $\sigma_\alpha$ so that Taylor expansion of $q_{a,\alpha}$ around the $q_{\tilde{s},\alpha}$ is valid.
Note that when plotted as $\Lambda_\sigma / \sigma$ (Inset to the second panel of Figure \ref{fig:LambdaEpsilonAndSigma}) the curves for small $\sigma$ values ($\le 0.1$) converge toward a {\em universal curve} which peaks around $\kappa \sim 3.05$.  Since the contribution of $\sigma_\alpha$ texture to $ \triangle_\rho$ is given in $\sigma \, \Lambda_\sigma ,$ this implies that the contribution to the shift becomes second order in $\sigma$. This is consistent with the comparison made earlier between the exact and the perturbative solutions in the case of a spherical pore\citep{Ryu:2009p753} and with the symmetry requirement, as the shift should remain independent of the sign of $\sigma.$

Figure \ref{fig:LambdaEpsilonSigmaLarge}  shows the factors that control $\delta \lambda_0$ for the symmetric ($\epsilon \ne 0, \sigma = 0$: top panel) and the asymmetric ($\epsilon = 0, \sigma > 0$: bottom panel) cases while the geometry is held fixed. In the symmetric case, the contributions for the positive and the negative $\epsilon$'s display strong asymmetry for large $\kappa$. Note that we had observed earlier that these tend to cancel out for small $\kappa$'s due to the condition Eq.\ref{eq:conditionforepsilon}.  For a large $\kappa$, it is no longer the case.
The second panel largely duplicates what we had found for the spherical pore.\citep{Ryu:2009p500}
The peculiar evolution of this factor from a peaky structure to a step-like shape as $\sigma\rightarrow 1$ can be understood if we examine the way profile of the eigenmode evolves as $\sigma$ increases. Figure \ref{fig:groundstate} contrasts the profile $\phi_0^0(x)$ of the slowest mode with $\sigma=0$, (shown with a broken curve in all panels) and $\phi_0(x)$ for finite $\sigma$ values (solid curves). The changes between $\phi_0^0$ and $\phi_0$ are shaded gray. The left column is for $\kappa=2$, right column with $\kappa = 32.$ For small $\kappa$ and $\sigma$, (top-left), the effect is generally a moderate shift in phase. As $\sigma$ increases, the wavelength tends to shrink. This is most pronounced in the large $\kappa-\sigma$ values (right-bottom panel). Note that in this limit, one has a large $\rho$ on one side, and vanishing $\rho$ on the other. Therefore, the system evolves from where the span $L$ of the pore matches the half-wavelength of $\phi_0^0$ (broken curve) to a highly asymmetric profile (solid curve in the bottom-right panel) in which $L$ equals the quarter-wavelength of $\phi_0$. Doubling of the length scale in $\phi_0$ leads to a decrease in $\lambda_0$ by a factor of 4, as indicated by convergence toward $0.75$ as $\sigma \rightarrow 1, \kappa \rightarrow \infty$ in the bottom panel of Figure \ref{fig:LambdaEpsilonSigmaLarge}. 

\section{Quadrature $\rho$ texture on a Rectangular pore}
\label{sec:quad2}
The classes of texture considered in the previous section capture basic aspects of $\rho$ and geometry in their entanglement with each other.  
However, it misses a subtle ingredient that may play an important role in media with nontrivial geometry. 
Recall how we showed that the first order $\epsilon_\alpha$ contribution to $\delta \lambda_{s,0}(0)$ vanishes. 
That arose from the constraint that $\bar \rho_\alpha$ is uniform in each plane. 
For more general situations, it is not necessarily so as the symmetry of $\delta \rho$ and the eigenmodes $\phi_p$ on the interface play an interesting role.  
\begin{figure}[h] 
   \centering
   \includegraphics[width=2.4in]{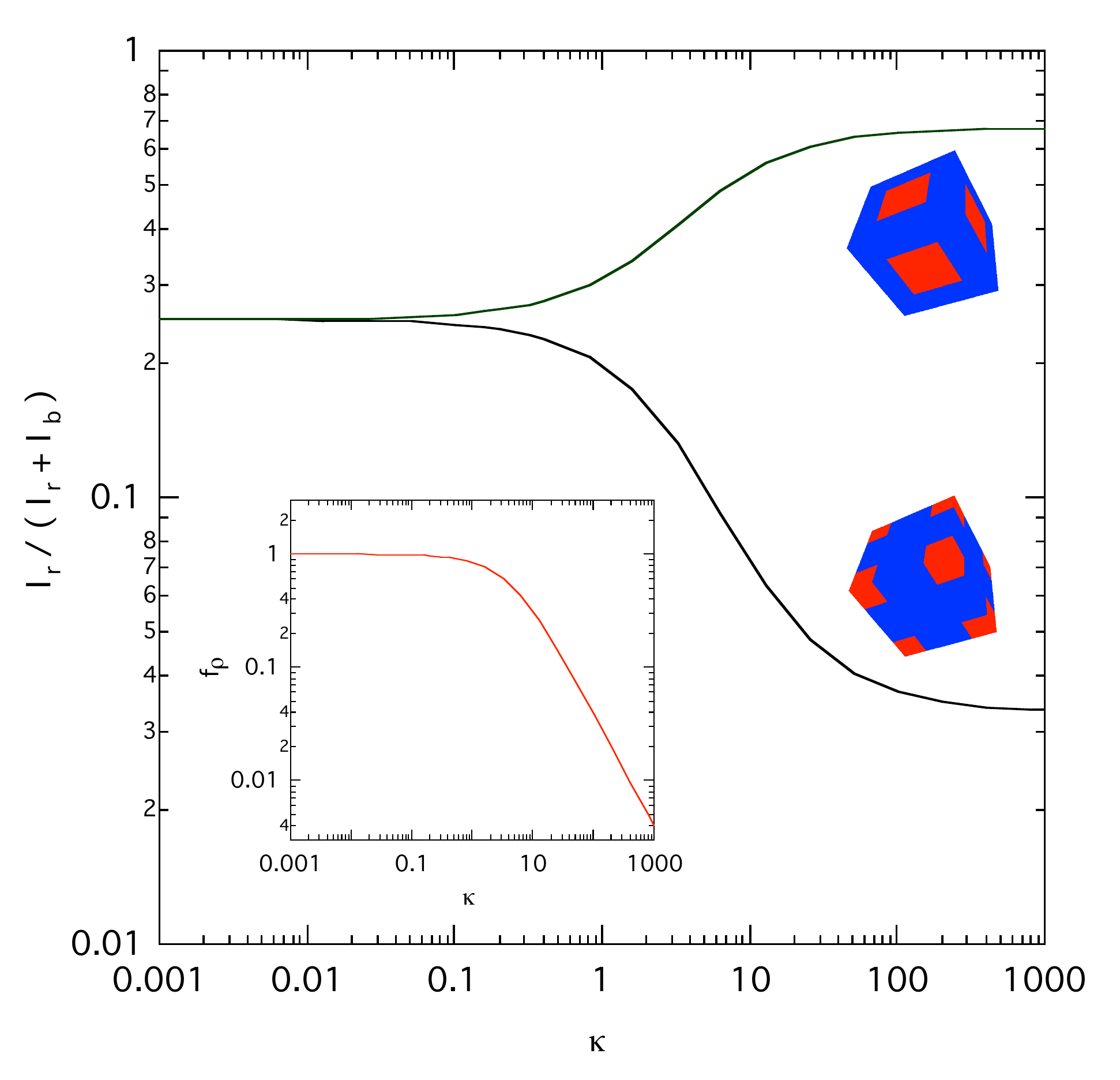}  \includegraphics[width=2.5in]{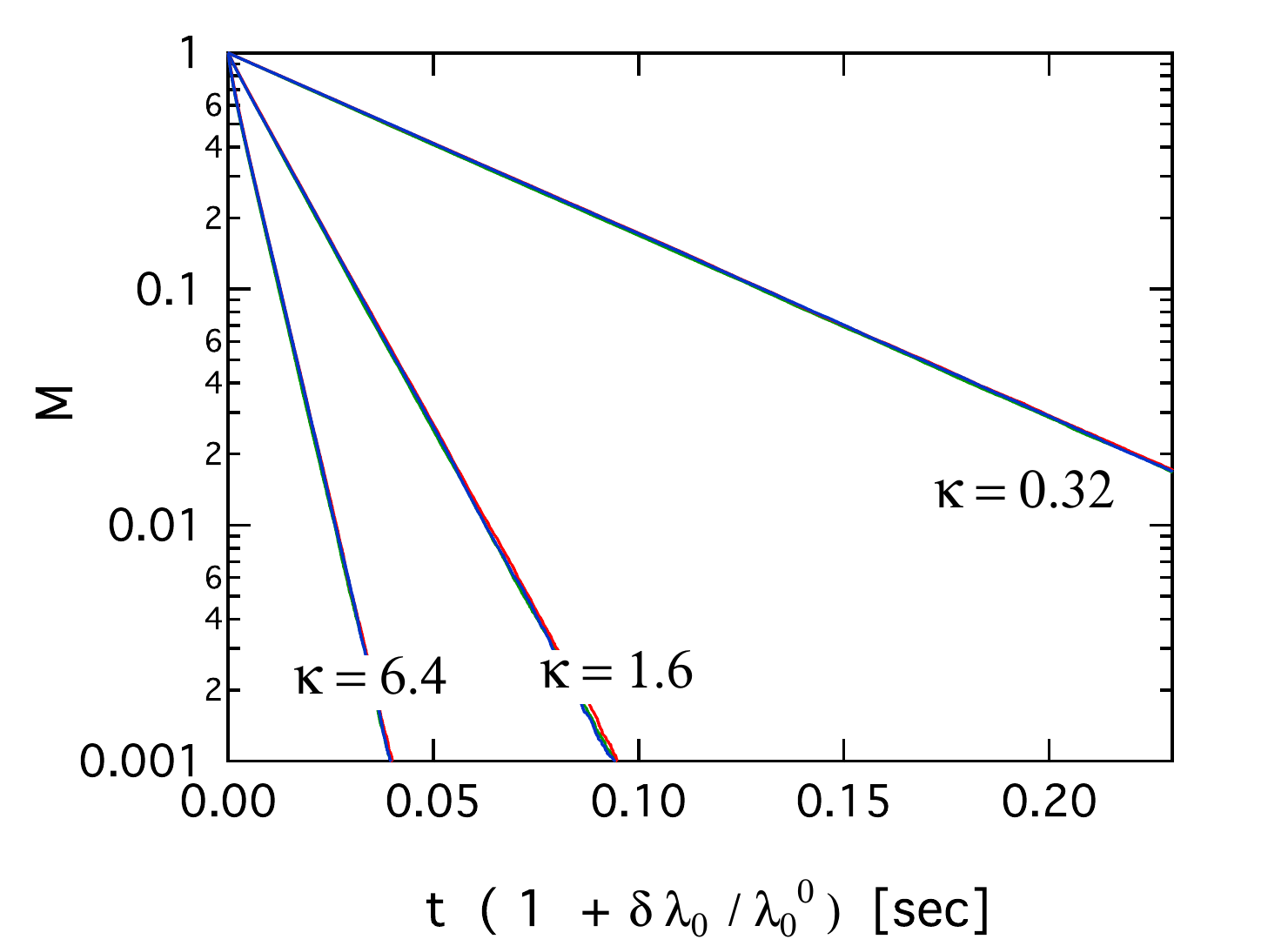} 
   \caption{\small Top panel: Fractional surface-integral factor of $(\phi_0^0)^2$ over the red area on which $\rho(\r)$ is enhanced. 
   The relative weights of $I_r$ and $I_b$ controls whether the rate increases or decreases with respect to the reference case. Their degree is further controlled by the $f_\rho=(1 + \rho_0 \ell_0 / D)^{-1}$ factor which is a purely geometrical parameter and is shown in the inset. 
 The second panel shows the curves of Figure \ref{fig:drhoCubeSimResults} with time scaled by the factors $(1 + \delta \lambda_0 / \lambda_0^0)$ evaluated using values of $f_\rho$ and $I_r$ above .  Total of nine curves are shown, three each (following the same color convention for each $\delta\rho$ texture for the three $\kappa$ values. The three curves collapse to a single curve for each $\kappa$.    }
   \label{fig:drhoCubeFractions}
\end{figure}
For the leading oder contribution to the shift, we showed:
\begin{equation}
\label{eq:firstorder}
\delta \lambda_0 = \oint_\Sigma d\sigma \phi_0^0 (\r)  \delta \rho (\r)  \phi_0^0 (\r) + {\cal O}(\delta \rho^2)
\end{equation}
with $\delta \rho$ constrained to satisfy
\begin{equation}
\label{eq:firstorder1}
\oint_\Sigma d\sigma \delta \rho (\r) = 0.
\end{equation}
In the case of a spherical pore, we had $\phi_0^0$ uniform across the interface, and therefore this first order contribution was shown to vanish.  It was further shown that the higher order contribution always acts to slow down the decay, i.e. $\lambda_0^0 >  \lambda_0$.
In a rectangular pore, it is no longer the case for a general $\delta \rho$ and we have a chance to observe the first order effect if $\delta \rho(\r)$ variation is chosen to have a non-trivial overlap with respect to $\phi_0^0(\r)$. It is further expected that  $\delta \lambda_0$ may become either positive or negative.

Let us demonstrate that  such a case is readily observable with a rectangular pore.  With a uniform $\rho_0$,  its slowest mode is readily obtained:
\begin{equation}
\label{eq:slowestmodecubeuniform}
\phi_0^0 (\r) =   \prod_{\alpha = 1}^{d_e} \sqrt{ \frac{1}{ \frac{L_\alpha}{2} (1 + \frac{\sin (k_{0,\alpha} L_\alpha)}{ k_{0,\alpha} L_\alpha } )} } \cos (k_{0,\alpha} r_\alpha)
\end{equation}
with each $r_\alpha \in [-L_\alpha/2, L_\alpha/2]$. 
The first panel of Fig.\ref{fig:groundstate} shows the relative strength of $\phi_0^0$ in a cubic pore for $\kappa_\alpha (= \rho_0 L_\alpha / D) = 1.6$.  The colormap is chosen to emphasize the depletion of $\phi_0^0$ near the eight corners of the cube. Note that within each interfacial plane, $S_\alpha$ with its surface normal $\hat{n}_\alpha$, $\phi_0^0$ has a local maximum at the center of the plane. The next three panels show the profile of $(\phi_0^0)^2$, as it appears in Eq.\ref{eq:firstorder}, in one of the interfacial planes for three values of $\kappa = 0.32, 1.6,$ and $6.4$ normalized with respect to its value at center of the plane.  
Now we consider three $\delta \rho$ textures as described in Figure \ref{fig:drhoCube} imposed a cube of size $L^3$. Two cases, one with enhanced $\rho$ strength in the square centered on each plane (designated as face-centered), another with enhancement in the eight corners of the interface (cornered) are compared with respect to the uniform $\rho_0$.  
For the uniform case, the rate is given by
\begin{equation}
\label{eq:lam00quad}
\small
\lambda_0^0 = \rho_0 \oint d\sigma \phi_0^0(\r)^2   ( 1 + \kappa \frac{\ell_0}{L})  = \rho_0 (I_r + I_b)  ( 1 + \kappa \frac{\ell_0}{L})
\end{equation}
the fractional shift, Eq.\ref{eq:firstorder}, becomes
\begin{eqnarray}
\label{eq:firstordershiftquad}
\small
\lefteqn{ 
\frac{\delta \lambda_0}{\lambda_0^0} = \frac{\oint_\Sigma d\sigma \phi_0^0 (\r) (\rho (\r) - \rho_0) \phi_0^0 (\r) }
{ \oint_\Sigma d\sigma \phi_0^0 (\r) \rho_0 \phi_0^0 (\r) } f_\rho  + {\cal O}(\delta \rho^2) 
} \\
&& \rightarrow  \Big(  (\frac{\rho_r}{\rho_0}-1) \frac{I_r}{I_r + I_b} +(\frac{\rho_b}{\rho_0}-1) \frac{I_b}{I_r + I_b} \Big) f_\rho  
\nonumber
\end{eqnarray}
where  $f_\rho \equiv 1/(1 + \rho_0 \ell_0 / D)$ and 
\begin{equation}
\label{eq:defineir}
I_{r} =\frac{L}{6}  \oint _{\Sigma_r} d\sigma  |\phi_0^0 (r)|^2
\end{equation}
is the surface integral restricted to the red part ($\Sigma_r$) of the interface in the Figure \ref{fig:drhoCube} 
and similarly for $I_{b}$. The particular choice we made as depicted in the figure leads to the average 
$ \rho_0 = \rho_b  \frac{3}{4} + \rho_r  \frac{1}{4}$ where $\rho_{r (b)}$ is the local $\rho$ value on the red (blue)-part, which takes up $1/4$ of each plane.

Figure \ref{fig:drhoCubeSimResults} shows the simulated relaxation curves obtained using the random walk method\citep{Ryu:2008p531} for the three cases in each panel. Three panels with progresively larger $\kappa$ values are shown. 
In these calculations, we further chose to have $\rho_r = 2 \rho_b$. (therefore $\rho_0 =  \frac{5}{4}\rho_b$)
The overall trend is that the slowest rate for the face centered pattern got faster, while the rate for the cornered texture got slower compared to the reference case with the uniform $\rho_0$. The trend becomes more pronouced for larger $\kappa$. The same can be observed in the $T_2$-distributions of the same data shown in the insets, although the amount of change may not look unambiguous for those experienced with the inherent uncertainty in such a representation. The trend in the corresponding time-domain curves is free from the inversion-related issue and is real. The broken curves attached to each time domain graph at early times indicate the predicted $\exp(-t/\tau_s)$ behavior where   
$\tau_s = \rho_0 S / V$. $1/\tau_s$ is also indicated in the inset as a short yellow bar. 

For these calculations, the first order contribution to the fractional shift of Eq.\ref{eq:firstordershiftquad} is shown to be   
\begin{equation}
\label{eq:quadshifts}
\frac{\delta \lambda_0}{\lambda_0^0} 
=  \Big(   \frac{4 I_r}{I_r + I_b} -1 \Big) \frac{f_\rho   }{5}.
\end{equation}
Note that in the Figure \ref{fig:drhoCubeFractions},  the values of $I_r/ (I_r + I_b) \ge 0.25$ for the centered texture, while it is $\le 0.25$ for the cornered texture. Thus the first order contribution decreases the rate for the cornered and does the opposite for the centered. 
Reading the values of $\frac{I_r}{I_r + I_b}$ and $f_\rho$ from Figure \ref{fig:drhoCubeFractions}, we obtain, for the cornered texture, $\delta \lambda_0 / \lambda_0^0 =  -0.014295,  -0.0465, -0.0542$ for $\kappa = 0.32, 1.6, 6.4$ respectively. 
For the face-centered texture, the rate is predicted to increase with $\frac{\delta \lambda_0}{ \lambda_0^0} =  0.0149, 0.0548, 0.08069$ for $\kappa = 0.32, 1.6, 6.4$.
This is in excellent agreement with the simulated results, as indicated in the last panel of Figure \ref{fig:drhoCubeFractions} where we show that the curves for different textures from Figure \ref{fig:drhoCubeSimResults} all collapse when plotted with respect to the time scaled by the 
$(1 + \delta \lambda_0/ \lambda_0^0)$, using the fractional shift values found above.
The agreement hardly leaves any room for the higher order perturbative contribution to make any significant addition. 

\begin{figure}[tb] 
   \centering
   \includegraphics[width =2.8in]{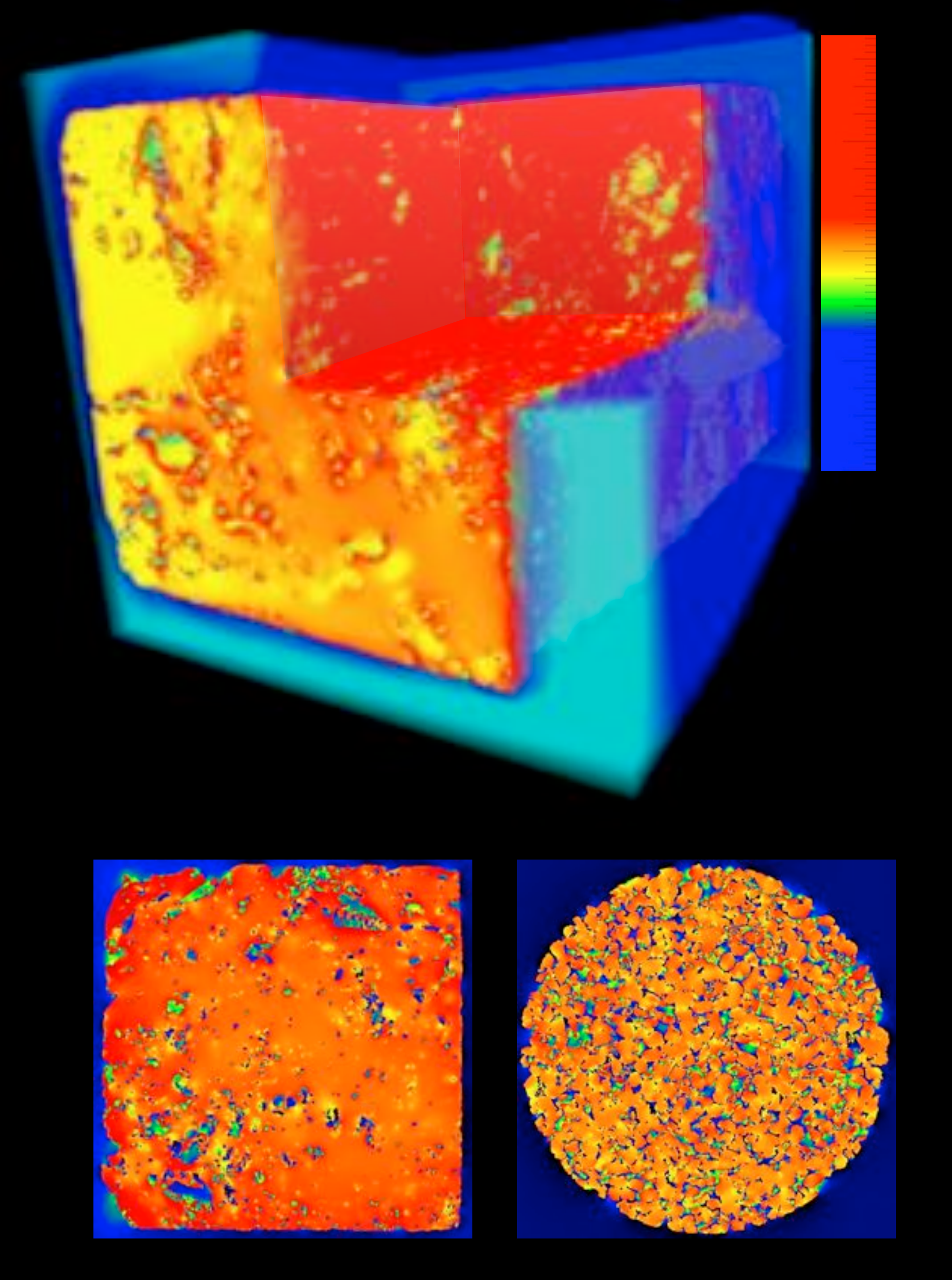} 
 \caption{\small Top panel: Internal field calculated to first order in $\triangle \chi$. The sample is a carbonate rock (packstone) of distorted rectangular shape with dimensions $1\times 1 \times 1.3 \, {\rm cm}^3$. The applied field is along the long axis of the sample.  The calculation was done for the entire tomogram volume $1.5\times 1.5 \times 1.3 \, {\rm cm}^3$ including the free space surrounding the rock, which is mirror-reflected and then periodically repeated. The colormap was adjusted for optimal contrast of the features. The part for the surrounding water is displayed with enhanced transparency to lessen obstruction of view. Part of the rock was cut out to reveal its inside. Bottom panels: cross-sectional cutouts of the internal field. The left panel is from the same sample as above. The right panel is from a Berea sandstone with cross-sectional dimension of $5.2^2 {\rm mm}^2$.}
   \label{fig:carbonatehz}
\end{figure}

\begin{figure}[tb] 
   \centering
   \includegraphics[width=2.4in]{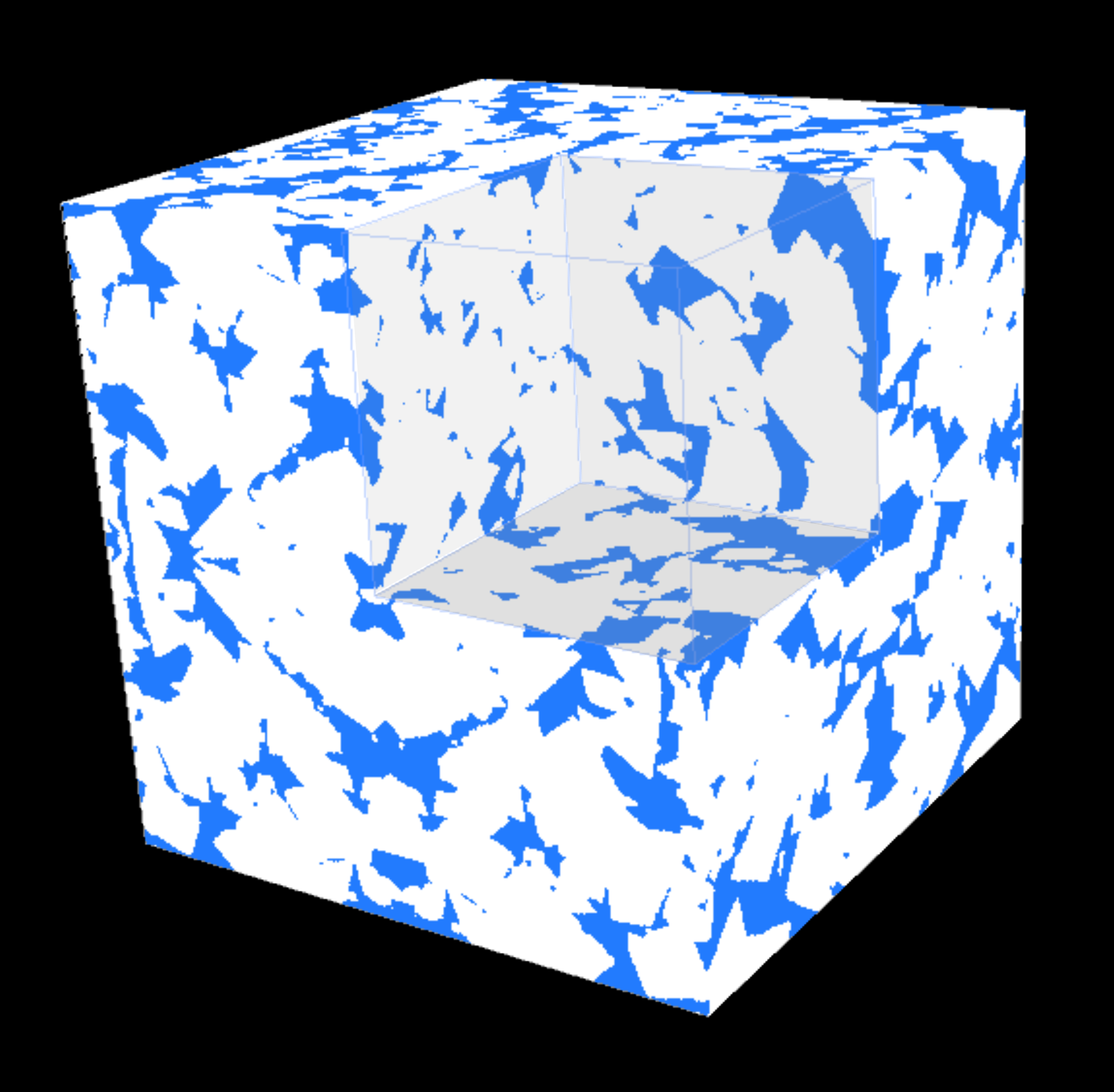} 
 \includegraphics[width =2.4in]{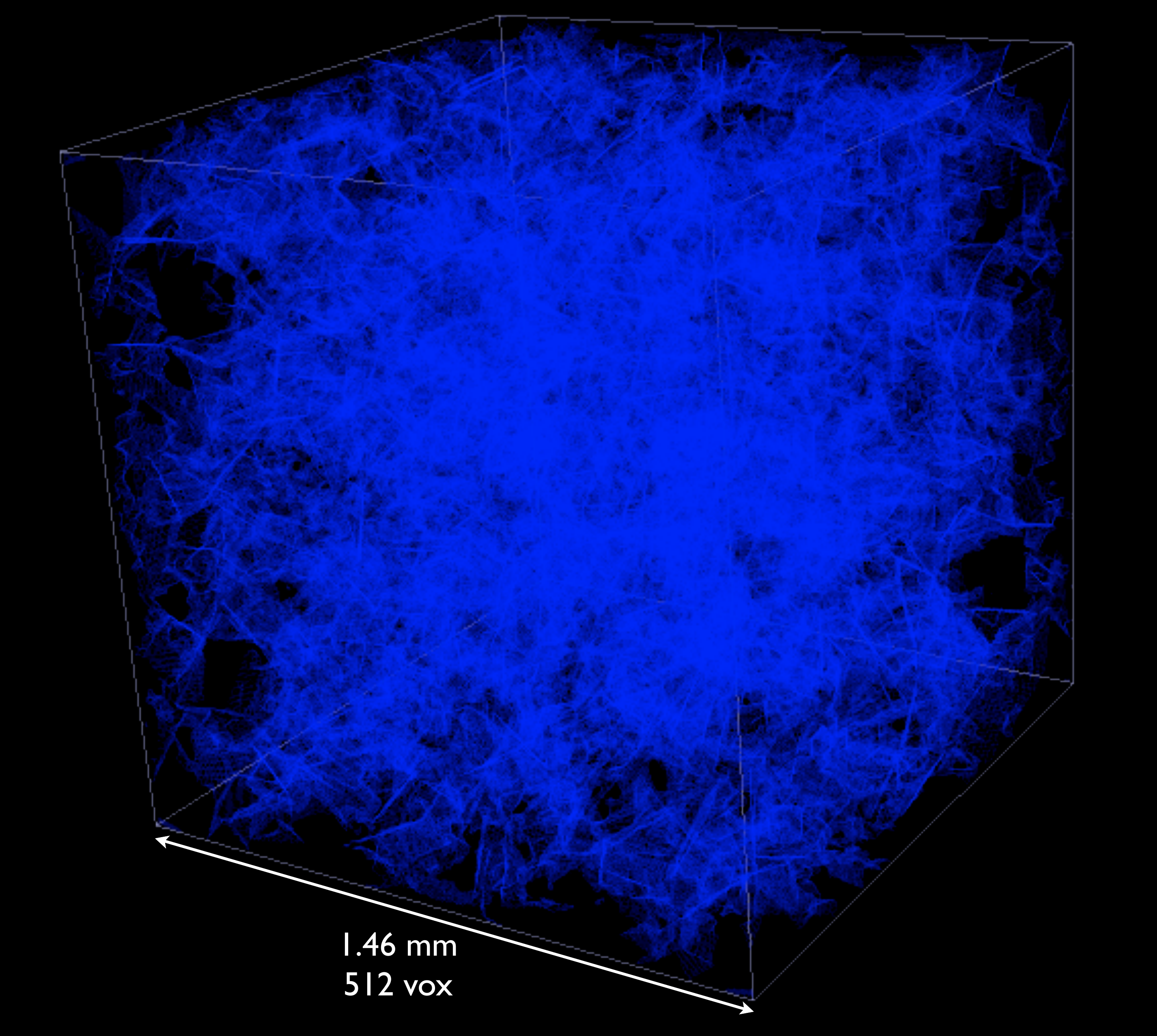} 
   \caption{\small $512^3$ volume ($1.46^3 {\rm mm}^3$) of a high resolution tomogram of a carbonate rock used in NMR simulations. Blue is the pore space, white represents the grains. The volume consists of mirror-reflected images of the seed volume of dimension $256^3$ as indicated by the shaded cutout. The lower panel shows the pore-grain interface only.}
   \label{fig:carbonatepore}
\end{figure}

\section{Internal field-like $\rho$ texture vs. Patched $\rho$ on a carbonate rock}
\label{sec:hz} 
So far, we have not addressed whether there should exist any correlation between the spatial profile of $\delta \rho(\r)$ and the underlying pore geometry in natural media. When it does, it would depend on the geological history of the formation; Extrinsic factors such as the  presence of strong magnetic minerals and their distribution may vary in a haphazard manner from one area to another, and one cannot expect to have a {\em generic} profile that covers them all. Systematic experimental analyses are emerging only recently with relevant details \citep{Keating:2007p846}.
In previous sections, we considered a few artificially imposed textures: In the first section, it was uniform in each interfacial plane, but allowed to vary from plane to plane. In the preceding section, patterns were imposed deliberately to control degrees of commensuration with the $\phi_0(\r)$. In our earlier work, \citep{Ryu:2008p531} we employed grain-specific assignments in a random beads pack as well as correlated-random noise (patchy) pattern with varying correlation lengths. In a spherical pore, both exact solution and numerical simulations were obtained \citep{Ryu:2009p500} for hemispherical assignment. Arns {\em et al} \citep{Arns:2006p581} used grain-assignment in numerical simulations on pores generated from tomograms.  Using random assignment at a voxel-level, Valfouskaya {\em et al} \citep{Valfouskaya:2006p549} observed negligible effect. 
Most of these observations can be understood within our theoretical framework yet there are sources of $\delta \rho$ that defy an easy categorization. As an example, we consider a case where the surface-relaxation may vary in strong registry with a aspect of pore geometry, yet it is not clear {\em a priori} whether it would work in its favor or against.  

The most widely accepted mechanism for the microscopic origin of $\rho$ is based on the engagement of a migrant proton spin with a surface-embedded paramagnetic ion spin. \citep{Korringa:1962p646, Brown:1961p856, Kleinberg:1996p811}
On a larger length scale, the pore matrix and its filling fluid have different magnetic susceptibilities (let us denote the difference as $\triangle \chi$), and this gives rise to an internal field ${\bf B} (\r)$. This leads to an inhomogeneous Larmor frequency, leading to extraneous dephasing (so-called secular relaxation) of transverse spins while it diffuses around during the time interval $\tau_E$
between the $\pi/2$ and $\pi$ pulses in a typical NMR echo measurement. 
Its effect has been studied for a constant gradient \citep{Sen:1999p865}, a parabolic field (i.e. linear field gradient) \citep{LeDoussal:1992p864} and also for a periodic  case. \citep{Bergman:1995p866}. 
A spherical pore and the adjacent spherical dipole source were considered by Valckenborg {\it et al} \citep{Valckenborg:2003p867}.
Gillis {\em et al} \citep{Gillis:2002p609} noted the {\em strong} gradient variation in the case of a single spherical source, pointing to more faithful account of the realistic field profile. Recently, Anand {\em et al} \citep{Anand:2007p741} considered the field produced by an arrangement of dipole sources placed on a spherical grain and considered qualitatively different regimes.

\begin{figure}[h] 
   \centering
 \includegraphics[width =2.8in]{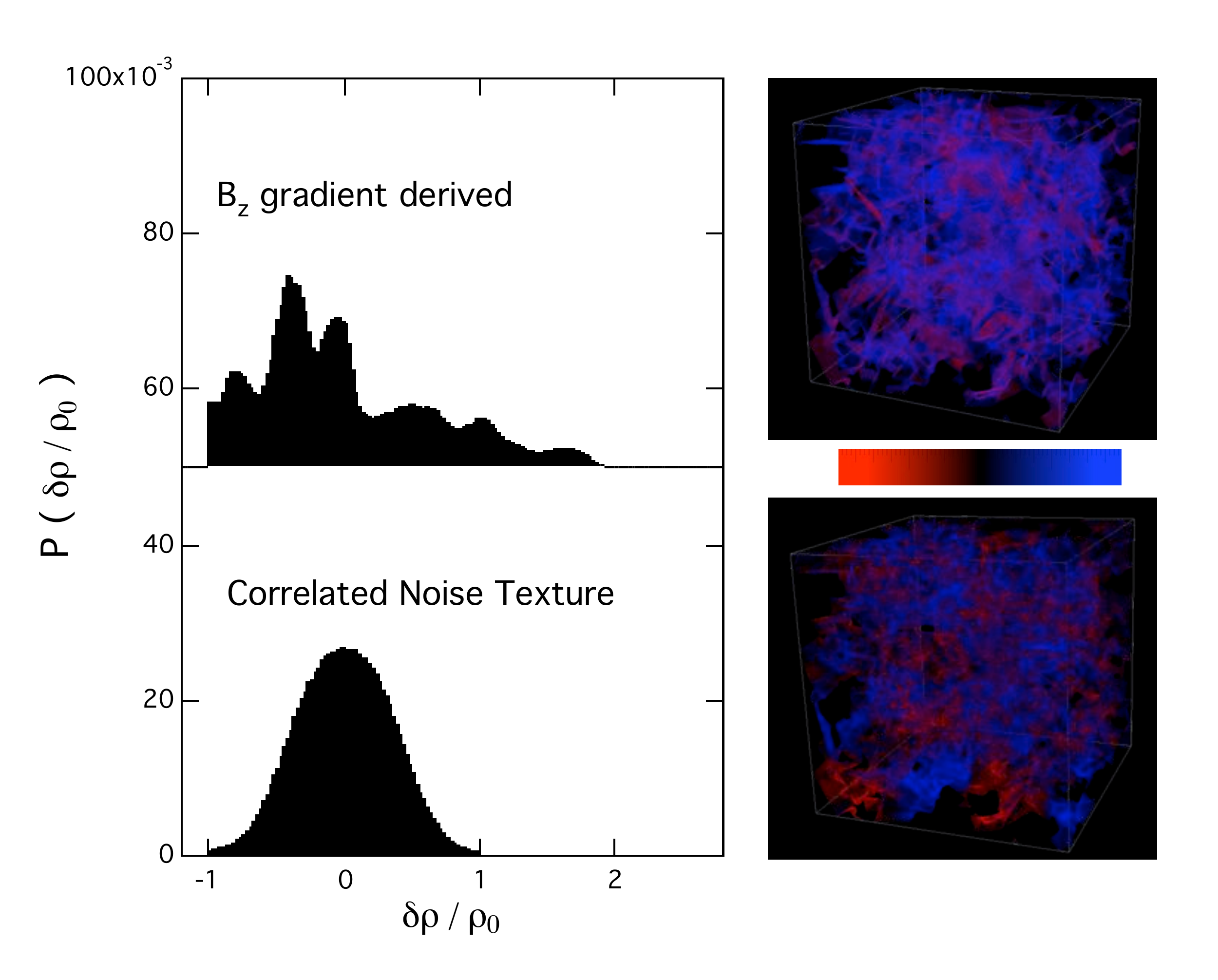} 
 \caption{\small Two $\delta \rho$ textures imposed on the interface of Figure \ref{fig:carbonatepore}. From numerically calculated internal field, $B_z(\r)$,  its local gradient strength $|\nabla B_z(\r)|$ is obtained  using the finite difference scheme, and $\delta \rho_B$ is derived from it (shown in the top panel along with its probability distribution on the interface). The bottom panel shows the texture generated using the correlated random noise sequence for comparison. In the latter, the pattern has a shorter correlation length compared to the former, although there are occasional large patch areas of enhanced $\rho$ (e.g. red zones in the bottom corner).  The rms-deviations for the distributions  are $0.65$ and $0.35$ respectively.}
   \label{fig:carbonatedrhocompare}
\end{figure}

\begin{figure}[h] 
   \centering
 \includegraphics[width =2.5in]{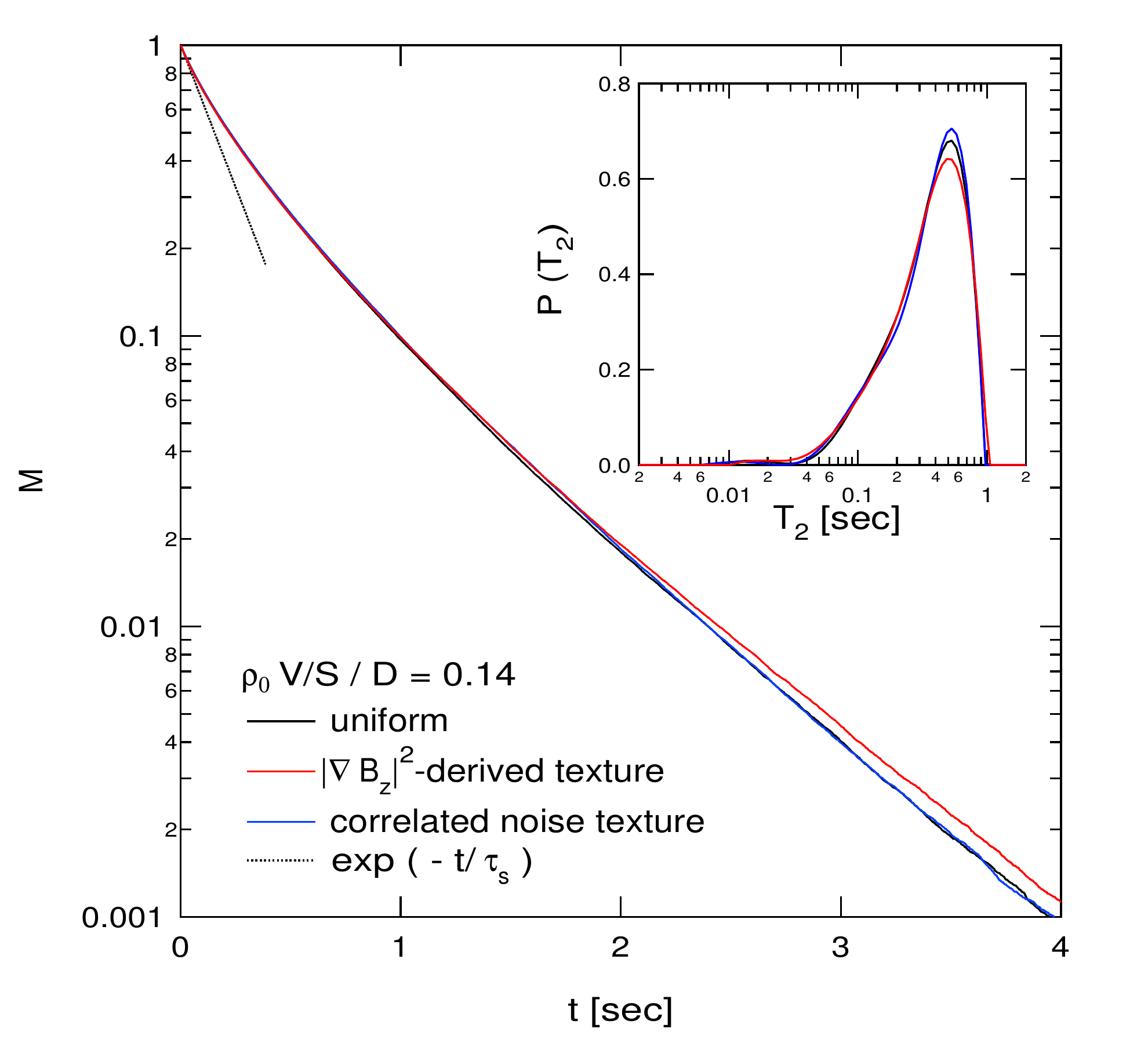} 
\includegraphics[width =2.5in]{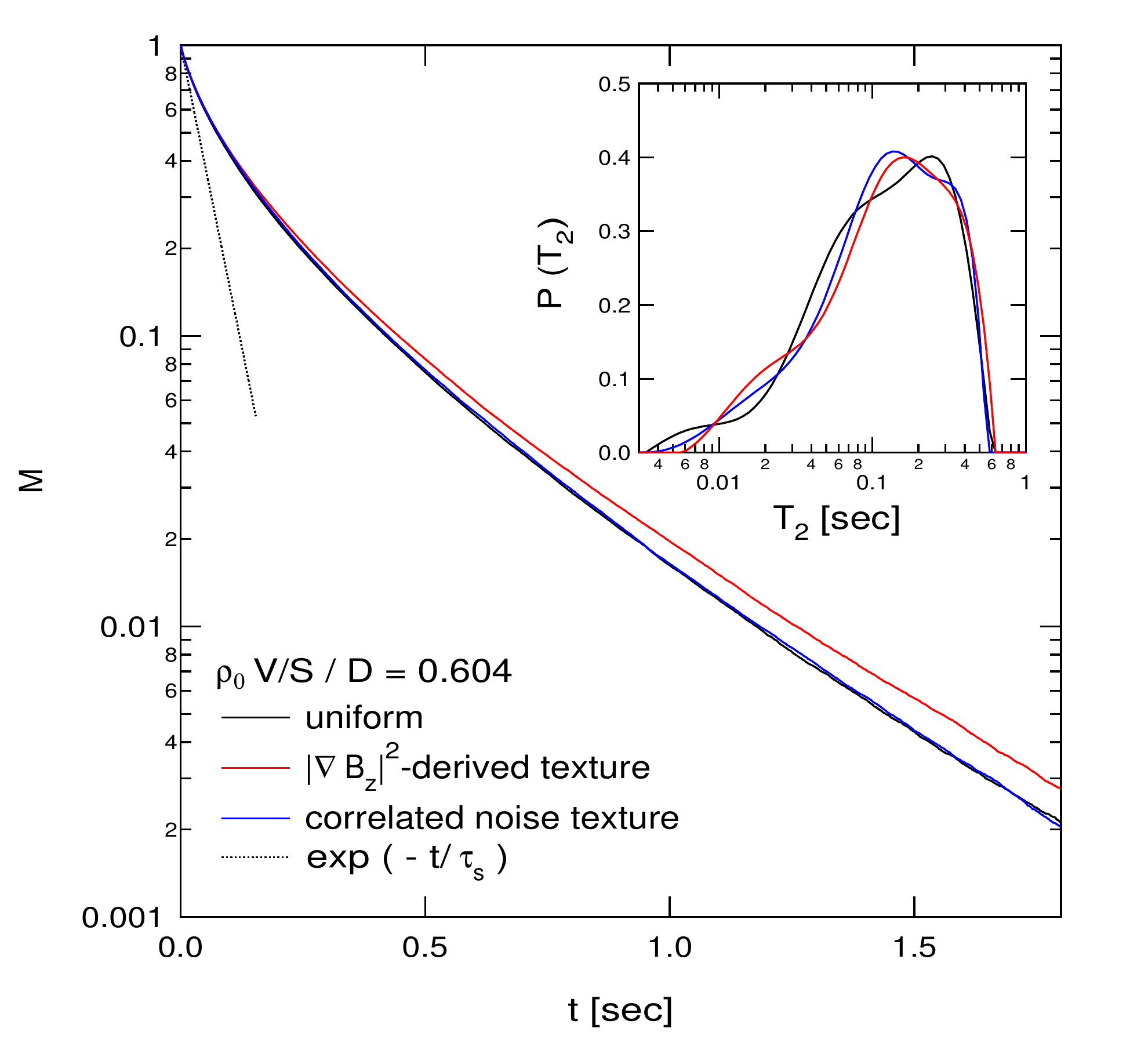} 
 \caption{\small $M$ vs. $t$ for the carbonate pore of Figure \ref{fig:carbonatepore} for different textures of $\rho$: uniform $\rho_0$ (black), $B_z$-gradient derived $\delta \rho_B$ (red), and correlated random noise generated texture (blue). Short broken lines indicate the predicted initial slope with $\tau_s^{-1} = \rho_0 S/V$. The insets show the $T_2$-distributions with the same color scheme. Top panel is for $\rho_0 = 40 \mu m/{\rm sec},$ bottom panel is for $\rho_0 = 170 \mu m/{\rm sec}$. In both cases, $V/S = 8.88 \mu m$, $D = 2500 \mu m^2/{\rm sec}$. Porosity of the sample was $0.23$.}
   \label{fig:carbonatedrhopt2}
\end{figure}
The inhomogeneous ${\bf B} (\r)$ in real rocks displays some aspects not explicitly captured in these studies. 
For a simple geometry, it is obvious that the internal field gradient strength depends strongly on the orientation of the interface. 
It is further noted that the field variation is strongly enhanced closely along the interface with the right orientation and rapid pore shape variation.\citep{Ryu:2001p705} Start with a field profile from an isolated spherical dipole of radius $a$, then extend it for a polydisperse glass beads pack as an example of complex pore matrix. From the exercise, we hypothesize that the interface is lined with a layer in which the innternal field gradient strength is pronounced with its order of magnitude given by 
$|\nabla B| \sim \triangle \chi B_0 / \epsilon$ where  $\epsilon$ is the typical length scale for the local curvature of  corrugated pore interface. $\epsilon$  may be a small fraction of the gain size and depends  on the roughness of the interface. 
On a coarser length scale,  this rapid variation of local field would be smeared out, as the local interface orientation with respect to the applied field would change rapidly. This coarser picture interpolates to what one would observe with a dense aggregation of dipoles on a  smooth spheres. 
A molecule diffusing in the zone over the echo spacing $\tau_E \le   \epsilon^2/D$, however,  would feel the field variation in a finer scale, and the effect on the accumulated phase of the transverse spin, will be equivalent to that of layer of thickness $\epsilon$ lining the part of the wall with an surface relaxivity 
\begin{equation}
\label{eq:definerhoh}
\rho_{B} \sim \epsilon (\frac{\gamma \triangle \chi B_0\tau_E}{\epsilon} )^2 D
\end{equation}
where we assume that $\gamma \triangle\chi B_0 \tau_E < 1$ to allow Carr-Purcell's classic argument to hold. (otherwise, it crosses over to the diffusion controlled regime with weakened dependence on $\tau_E$ but at values of $\rho_B$ higher than estimated below.)
For $B_0 = 500 G$, $D = 2500 \mu m/{\rm sec}^2$, we get 
$4.46\times 10^{11} \frac{\triangle\chi^2 \tau_E^2}{\epsilon}$ ($\tau_E$ in msec, $\triangle \chi$ in emu,  $\epsilon$ in $\mu$m).
For echo spacing of $1$ msec, a quartz sandstone with $\triangle \chi \sim 4 \times 10^{-6}$ and $\epsilon = 50 \mu {\rm m}$ would give
$\rho_B \sim  0.1$ ($\mu {\rm m / s}$).  For matrix with finer length scales $\epsilon \le 10 \mu{\rm m}$, and moderate susceptibility contrast $\triangle \chi \sim 10^{-4}$ (such as realized experimentally via coating of sand grains \citep{Keating:2007p846}), $\rho_B \sim 100 - 1000 \mu{\rm m/s}$ is reached. 
Whether this {\em secular} mechanism of relaxation overwhelms that due to embedded paramagnetic impurities is a question that should be asked for each sample ($\triangle \chi, \epsilon, D$) and the experimental setup ($\tau_E, B_0$), looking for its signature in its dependence on varying $\tau_E$. 
Having shown that it is a viable source of an apparent $\rho_B$ within an experimentally plausible range, we may view the spatial variation inherent in $\triangle \chi B_0 / \epsilon$ as $\delta \rho_B (\r)$, and ask the same question as to its impact on $\lambda_0$.

Detailed analysis of ${\bf B}(\r)$ in a large 3D porous rock is beyond the scope of this paper and will be reported elsewhere. Here we briefly summarize how we derive the pseudo-$\delta\rho$ texture using the technique we recently developed for the internal field and its gradients. The method has been tested for a variety of porous media including a packed cylinders for which direct experimental imaging was also done for comparision.\citep{Cho:2009p776}
Figure \ref{fig:carbonatehz} shows examples of such calculation applied to a large 3D tomogram of carbonate and sandstones.  The top panel shows the full volumetric rendition of the internal field. The bottom panels show cross-section of full 3D calculations for the carbonate and the Berea sandstone.
To minimize the boundary effect, the volume  $(1.5\times1.5\times1.3 {\rm cm}^3)$ including the surrounding fluid zone was also included in the calculation. It was further enlarged by mirror reflection images in all three directions. Then the mirror-imaged volume was periodically repeated via the FFT algorithm used in the calculation of the internal field components $B_\alpha/ (\triangle \chi B_0)$. The field gradient strength was calculated from the finite difference evaluation of $(\sum_\beta |\nabla_\beta  B_\alpha|^2)^{0.5}$.  

Figure \ref{fig:carbonatepore} shows the carbonate pore structure used for simulating the NMR response using the internal field-derived $\rho_B(\r)$ texture as the sole relaxation  mechanism. To avoid artificial blockage on the bounding surface, a sub-volume of $256^3$ voxels was taken deep inside the tomogram\citep{Sheppard:2004p795}, then mirrored in all directions. Then NMR simulation was performed \citep{Ryu:2008p531} with up to $0.8\times 10^6$ random walkers subject to the stochastic dephasing at the wall under the periodic boundary condition.
From the internal field $B_z$ and its gradient, we generate the following pseudo- texture:
\begin{equation}
\label{eq:hztexture}
\rho_B (\r) = \rho_{0} (1 +\frac{ |\nabla B_z (\r)|^2 - < |\nabla B_z|^2> }{< |\nabla B_z|^2>})
\end{equation}
where $\rho_{0}$ now corresponds to the effective relaxivity (Eq. \ref{eq:definerhoh}) arising from a uniform gradient strength with the average $< |\nabla B_z|^2>$. 
The probability distribution is shown in the upper panel of Figure \ref{fig:carbonatedrhocompare}. 
For comparison, we also applied the correlated-random noise sequence\citep{Ryu:2008p531} to generate a patchy $\delta \rho$ texture. 
Its distribution (lower panel of Figure \ref{fig:carbonatedrhocompare}) is symmetric and has about the half the rms-deviation as that of $\delta \rho_B$. 

The right panels show how the distribution looks like on the pore-grain interface. Note that the $|\nabla B_z|^2$-derived $\delta\rho_B$ shows a pronounced {\em planar} correlation, reflecting its sensitivity to the local orientation of the surface with respect to the field direction. It also has a skewed distribution. The carbonate used in this calculation has a pore structure somewhat more complex than a sandstone, but the grains have relatively smooth surface, therefore we do not seem to have strong variation of $|\nabla B_z|^2$ on sub-grain scales.  The correlated random noise texture has variations on relatively finer length scales. 
Given that there is a visibly pronounced correlation in $\delta \rho_B$ on the granular length scale, our previous findings would indicate that it might have a better chance of having an impact than the correlated random-case. They also have contrasting skewness in their distributions. In the latter case, unless a subtle effect such as the case with quadrature patterns considered above, one may expect a gross cancellation effect.

The results of simulation for two different values of $\rho_0 \frac{V}{S} / D = 0.14, 0.60$ are shown in Figure \ref{fig:carbonatedrhopt2}. 
Note that in both cases, the curves all have much pronounced multi-exponential characteristics compared to the curves we have for a simple cubic pore (Figure \ref{fig:drhoCubeSimResults}). This is understandable given the much more complex pore goemetry, yet a naive application of the $\kappa$-criterion would have one to expect a more single-exponential like relaxation. In all cases, the initial decay  faithfully follow the exponential $\exp(-t \rho_0 S / V)$ behavior (shown as broken lines) as predicted, with and without $\delta \rho(\r)$. Note that the range of agreement, however, becomes impractically narrow as $\kappa$ increases. 
Finally, we observe that the spatial variation of $\rho$, when it is prescribed to mimic the internal field gradient brings about a visible shift in the slowest rate, while $\rho$ with the correlated-random noise pattern hardly impacted the relaxation behavior with respect to the uniform case. 
The $T_2$ distributions, shown in the inset, display a hint of minor changes for larger $\kappa$, yet it is negligible for both cases given the inherent uncertainty in the inversion process. 
For the log interpretation, this implies that the explanation for the spread and shape of the $T_2$ distribution as shown in the figure or as one might have encountered in an NMR logging operation should be sought in the geometrical features of the pore matrix rather than in $\delta \rho (\r)$, as long as there is no compelling reason to assume an unusually strong magnetic perturbation. On the other hand, high quality laboratory data taken with controlled variations in the echo spacing, when analyzed in the time domain, may display the trend we observe numerically. Our findings suggest an interesting possibility to probe the internal field induced relaxation using artificial bead/grain packs in which one can vary the smoothness and anisotropy of individual grains and their arrangement.

\section{Conclusions}
Using the rectangular pore geometry, we developed an analytical method to systematically probe the intertwined effect of its geometrical and $\rho$ textural variations. The result yields the fractional changes in the slowest relaxation rate with respect to that of the uniform $\rho$ case for various values of $\kappa$ and $\sigma$ which control the geometry and $\rho-$texture respectively. This framework provides a bound for the uncertainty in the NMR log interpretation for complex formations. 
We  identified the relative symmetry of the slowest eigenmode and the interfacial pattern of $\delta \rho$ as factors that control the impact of $\delta \rho$ variation, and demonstrated the mechanism using a face centered and corner centered $\delta \rho$ in a cubic pore.  While it is plausible that the deposit pattern of paramagnetic ions may be incommensurate with the underlying pore geometry, it is not unthinkable that some correlation exist due to the diagenesis process constrained by the complex pore geometry. 
Currently, there is not empirical data with enough details to further narrow down on the $\rho$ texture. Experimental systems using coated glass bead packs may be designed to shed further light on the issue. Colloidal suspension with controllable interfacial properties is another system where our finding may prove useful. In real rocks, detailed mapping of magnetic profile of the pore-grain interface on small enough length scales would be invaluable for further progress. 

In the absence of such detailed empirical constraint in a typical NMR logging operation, we use as guidance simulation results based on artificially made $\delta \rho$ textures imposed on a tomogram-derived pore structure. Prevously, we demonstrated that the salient features observed on the simple closed-pore or bead packs remain in tact for a sandstone. There, we made comparisons among textures of $\delta \rho$, quasi randomly generated, with different degrees of spatial correlation. In the current work, we further made comparisons of a different nature, between one generated from the internal field gradient, and the other by the random noise sequence. The result suggests that gross cancelation due to symmetric grounds or short correlation lengths of the texture seem to mitigate the impact of $\delta \rho$ in a typical natural media, while stronger spatial correlation as in the internal field gradient profile, tends to act against such a trend leading to a weak, but observable shift.

The phase-space is too vast for drawing any conclusive generalization out of these numerical results alone. Still, in view of analytical results we elaborated in this work, it seems reasonable to conclude that  the effect of $\delta \rho$ on logging application is secondary only to the more significant effect of pore geometry variation as far as the dominant features of the $T_2$ distribution are concerned. The fractional shift and its inferred bounds (Figs.\ref{fig:LambdaL}-\ref{fig:LambdaEpsilonSigmaLarge}), when translated into uncertainties in the pore length scale, provide useful guidance when there is not enough information to determine the relative significance of the geometrical and the lithological variation in a given formation. 
While its impact seems minimal in the overall $T_2$ distribution of the NMR logging data, 
it is worth pointing out that the transport properties tend to be controlled by features of the matrix which are manifest only in a narrow range of scales.  The effect of internal field gradient, for having close relationship with pore geometry variations and also for inducing visible changes in NMR response, may be utilized toward such purpose once we gain better understanding and control over its impact. Therefore, in our effort to refine the NMR logging as a {\em more effective} permeability probe, detailed aspects of $\delta \rho$ and its impact on NMR response should merit further investigations.  
\section{ACKNOWLEDGEMENTS}
The author would like to acknowledge helpful discussions with D. Johnson (Schlumberger Doll Research). He also thanks A. Kayser, R. Wood (former colleagues at Schlumberger), J. Goebbels (The Federal Institute for Materials Research and Testing, Germany), and M. Knackstedt (Australian National University) for providing tomograms used for simulations in this work.

\bibliographystyle{spwla}
\bibliography{exactsolrect,exactsolrect_misc}

\vskip 0.1in
\section{ABOUT THE AUTHOR}
{\bf Seungoh Ryu} has a B.S. and M.S. degrees in Physics from  Seoul National University, South Korea, and a PhD in Applied Physics from Stanford University, USA. Before joining Schlumberger Doll Research, he worked on various aspects of photoelectron spectroscopy, high temperature superconductivity, vortex lattice dynamics, nonlinear dynamics of Josephson-junction arrays and related numerical simulations techniques. At SDR, his research focus has been on the relationship between the pore geometry of rocks and diffusive/transport properties of their filling fluid. His other research interest includes modeling of microfluidic sensors, parallel computing, and soft condensed matter physics. 

\end{document}